\documentclass[aps,prd,10pt,onecolumn,nofootinbib,showpacs,showkeys,superscriptaddress]{revtex4-2}

\usepackage[T1]{fontenc}
\usepackage[utf8]{inputenc}
\usepackage[english]{babel}
\usepackage{lmodern}

\usepackage{amsmath,amsthm,amsfonts,amssymb,bm}
\usepackage{accents}
\usepackage{mathrsfs}

\usepackage{graphicx}
\usepackage[table]{xcolor}
\usepackage{booktabs}
\usepackage{multirow}
\usepackage{dcolumn}
\usepackage{geometry}
\usepackage{natbib}
\usepackage{url}
\usepackage{textcase}
\usepackage{hyperref}
\usepackage{orcidlink}
\usepackage{silence}

\hypersetup{
  colorlinks=true,
  breaklinks=true,
  linkcolor=blue,
  citecolor=blue,
  urlcolor=blue
}

\allowdisplaybreaks

\makeatletter

\renewcommand*{\p@subsection}{}

\renewcommand*{\p@subsubsection}{}
\makeatother

\begin{document}
\title{Post-Newtonian N-Body Dynamics in
Extended Theories of Gravity}

\author{Antonio Tedesco\,\orcidlink{0000-0001-7414-4339}}
\email{antonio.tedesco.dt12@gmail.com}
\email{antonio.tedesco@uniroma1.it}
\email{antonio.tedesco@cref.it}
\affiliation{Dipartimento di Fisica, Sapienza Universit\`a di Roma, Piazzale Aldo Moro 5, 00185 Roma, Italy}
\affiliation{Enrico Fermi Research Center (CREF), Via Panisperna 89A, 00184 Roma, Italy}

\begin{abstract}
\fontsize{8.95pt}{11pt}\selectfont
We derive the complete first post-Newtonian (1PN) Lagrangian and corresponding equations of motion for the relativistic $N$-body system in Scalar-Tensor-Fourth-Order Gravity (STFOG), including the Non-Commutative Spectral Geometry (NCSG) sector as a special case. In the regime $\Phi \sim \Psi$ ($\gamma \sim 1$), the linearized fourth-order field equations are solved in the Standard Post-Newtonian gauge, and the variational Lagrangian is built directly from the point-particle action. The resulting dynamics is governed by three Yukawa functions $\zeta$, $\mathcal{W}$ and $\Xi$, which encode the scalar, gravitomagnetic and three-body sectors and depend on the effective masses $(m_R,m_Y,m_\phi)$ of the additional propagating modes. In this context, we show that the nonlinear metric component ${}^{(4)}\!g_{00}$ plays no role at 1PN level. The 1PN orbital motion of the above extended theories is thus obtained in closed form, and the Einstein--Infeld--Hoffmann equations are recovered in the corresponding general-relativistic limit. The formalism provides a common framework for relativistic celestial mechanics and applies to Solar System, binary pulsars such as PSR J0737-3039, Galactic-center stellar orbits and triple systems.
\end{abstract}

\date{July 12, 2026}
	
\pacs{04.25.-g; 04.25.Nx; 04.40.Nr}
	
\keywords{Extended theories of gravity; relativistic dynamics; tests of gravity.}

\maketitle
\begin{center}
\textit{Published in The European Physical Journal C, \textbf{86}, 885 (2026)}
\end{center}

\section{Introduction}\label{sec:introduction}
The Universe is dominated by the two dark components customarily identified as dark matter and dark energy. A broad body of observational evidence \cite{riess,ast,clo,spe,Carroll,sahini}, from galactic rotation curves to the cosmic microwave background and distant supernovae, supports their existence, but their fundamental nature remains elusive. These long-standing issues have kept alive a substantial interest in curvature-based extensions of General Relativity (GR), whose first systematic formulation dates back to Hermann Weyl \cite{Weyl1918RIG} in 1918, and in modified gravity as a complementary route to addressing such problems in a variety of astrophysical contexts. Among these proposals, the Extended Theories of Gravity (ETG) \cite{CapozzielloFaraoni,ETG} generalize the Einstein--Hilbert action by replacing the Ricci scalar $R$ with a more general function of curvature invariants, possibly in the presence of a non-minimally coupled scalar field. In four dimensions, the most general curvature-based class is Scalar-Tensor-Fourth-Order Gravity (STFOG) \cite{ETG,FOG_CGL}, with Lagrangian density $\mathcal{L} = f(R, R_{\mu\nu}R^{\mu\nu}, \phi) + \omega(\phi)\,\phi_{;\alpha}\phi^{;\alpha}$; it includes $f(R)$-gravity \cite{DeFelice:2010aj,Nojiri:2010wj}, scalar-tensor theories and the Non-Commutative Spectral Geometry (NCSG) model \cite{connes_1,ccm,mairi2012} as particular cases. In the weak-field limit the corresponding linearized field equations are of fourth order and the gravitational potentials pick up Yukawa-like corrections, whose effective masses are related to the extra degrees of freedom \cite{PechlanerSexl1966,stelle,PRD1,FOG_CGL}. 
While the Newtonian and post-Newtonian limits of these theories have been investigated for test-particle dynamics and several complementary post-Newtonian analyses for $f(R)$ gravity already exist \cite{Clifton2008,Lambiase:2013dai,Lambiase:2015yia,Berry:2011pb,CapolupoLambiaseTedesco,NaefJetzer2010}, a 1PN $N$-body dynamics for the full STFOG class has not been explicitly constructed so far. Filling this gap is the aim of the present paper. We derive both the relativistic post-Newtonian Lagrangian and the complete Euler--Lagrange equations for $N$ interacting bodies, extending the Einstein--Infeld--Hoffmann (EIH) dynamics \cite{EIH} to the regular STFOG branch; for systematic accounts of relativistic celestial mechanics in GR we refer to the monographs \cite{Brumberg,BrumbergBook1991,Kopeikin2011,Soffel}, to the collections \cite{KopeikinFrontiers1,KopeikinFrontiers2}, and to the relativistic $N$-body framework of Damour, Soffel and Xu \cite{DamourSoffelXu}. At 1PN order, the orbital dynamics is described by coupled timelike worldlines moving in the effective metric generated by the remaining bodies, with all beyond-GR corrections encoded in Yukawa functions associated with the additional scalar and tensor modes.

A key ingredient of our approach is the use of the variational principle for the field action \cite{Infeld,Brumberg2} directly at post-Newtonian level. Within this scheme one can check that the total 1PN Lagrangian does not depend on the nonlinear metric component ${}^{(4)}\!g_{00}$. In the Einstein--Hilbert sector this is the usual Brumberg cancellation between the field and matter contributions, while in the genuinely higher-curvature sector ${}^{(4)}\!g_{00}$ enters beyond the retained 1PN order. The property originally established in GR by Brumberg's conjecture \cite{Brumberg2,Brumberg3} is thus preserved in STFOG throughout the screened regime $\Phi\sim\Psi$, i.e. once the observationally well-supported hypothesis $\gamma\sim1$ \cite{Cassini,Will2014,71,72,73,74} is assumed. The final result is a closed set of 1PN $N$-body equations for STFOG, from which the $f(R)$, $f(R,\phi)$, $f(R,R_{\alpha\beta}R^{\alpha\beta})$, and NCSG sectors follow by specialization.
The paper is organized as follows. Section~\ref{sec:field-equations} presents the field equations and the post-Newtonian formalism, fixing the Standard Post-Newtonian (SPN) gauge in the subsequent $N$-body construction. Section~\ref{sec:brumberg} is devoted to the Brumberg conjecture, the Eddington--Robertson expansion and the relevant observational constraints. In Section~\ref{sec:linearized_field_equations} the linearized field equations are solved in the SPN gauge through the Green-function method. Section~\ref{sec:relativistic_dynamics} derives the $N$-body Lagrangian and the corresponding complete 1PN equations of motion for STFOG, with NCSG as a special case. A short discussion closes the paper in Section~\ref{sec:conclusions}.

\section{A general class for Extended Theories of Gravity}\label{sec:field-equations}

\subsection{STFOG action and field equations}\label{sec:STFOG}

Scalar-Tensor-Fourth-Order Gravity (STFOG) is the most general fourth-order curvature-based theory, including $f(R)$, $f(R,\phi)$, and $f(R,R_{\mu\nu}R^{\mu\nu})$ theories as special cases. Therefore, we consider it a representative class of the Extended Theories of Gravity. We adopt the signature $(+,-,-,-)$ and the convention $R_{\mu\nu} = {R^\sigma}_{\mu\sigma\nu}$. The STFOG action reads \cite{ETG,FOG_CGL}
\begin{equation}\label{FOGaction}
\mathcal{S}\,=\,\int d^{4}x\sqrt{-g}\,\biggl[f(R,R_{\mu\nu}R^{\mu\nu},\phi)+\omega(\phi)\,\phi_{;\alpha}\phi^{;\alpha}+2\,\mathcal{X}\,\mathcal{L}_m\biggr]\,,
\end{equation}
where $\mathcal{X}=8\pi G/c^4$ and $g=\text{det}\, g_{\mu\nu}$, $f$ is a generic function of the Ricci scalar $R$, of the curvature invariant denoted by $Y = R_{\mu\nu}R^{\mu\nu}$, and of the scalar field $\phi$; $\omega(\phi)$ is a generic coupling function, while $\mathcal{L}_m$ is the minimally coupled ordinary matter Lagrangian density.
The action \eqref{FOGaction} encompasses, as particular subclasses, $f(R)$-gravity (obtained by setting $f = f(R)$, $\phi = 0$), $f(R,\phi)$-gravity (setting $f_Y = 0$), $f(R, R_{\mu\nu}R^{\mu\nu})$-gravity (setting $\phi = 0$), and Brans-Dicke-like scalar-tensor models. We note that the Kretschmann invariant $R_{\mu\nu\rho\sigma}R^{\mu\nu\rho\sigma}$ can be discarded in four dimensions by virtue of the Gauss-Bonnet topological identity
\begin{equation}\label{GaussBonnet}
\mathscr{G} = R_{\mu\nu\rho\sigma}R^{\mu\nu\rho\sigma} - 4\,R_{\mu\nu}R^{\mu\nu} + R^2\,,
\end{equation}
whose integral reduces to a boundary term and does not contribute to the field equations \cite{CapozzielloFaraoni,stelle}. Consequently, any Lagrangian density of the form $\mathcal{L} = a\,R^2 + b\,R_{\mu\nu}R^{\mu\nu} + c\,R_{\mu\nu\rho\sigma}R^{\mu\nu\rho\sigma}$ can be reduced, up to a surface term, to an equivalent density $\mathcal{L} = \tilde{a}\,R^2 + \tilde{b}\,R_{\mu\nu}R^{\mu\nu}$, so that the STFOG action \eqref{FOGaction} provides the most general curvature-based extension of GR in four dimensions that also includes a scalar field.

Applying the variational principle to \eqref{FOGaction} with respect to $g_{\mu\nu}$ and $\phi$ yields the field equations \cite{ETG,FOG_CGL}
\begin{eqnarray}\label{fieldequationSTFOG}
&&f_R\,R_{\mu\nu} - \frac{f + \omega(\phi)\,\phi_{;\alpha}\phi^{;\alpha}}{2}\,g_{\mu\nu} - f_{R;\mu\nu} + g_{\mu\nu}\,\Box f_R + 2\,f_Y\,{R_\mu}^{\alpha}\,R_{\alpha\nu}
\nonumber\\[4pt]
&&\quad - 2\bigl[f_Y\,{R^\alpha}_{(\mu}\bigr]_{;\nu)\alpha} + \Box\bigl[f_Y\,R_{\mu\nu}\bigr] + \bigl[f_Y\,R_{\alpha\beta}\bigr]^{;\alpha\beta}\,g_{\mu\nu} + \omega(\phi)\,\phi_{;\mu}\,\phi_{;\nu}\,=\,\mathcal{X}\,T_{\mu\nu}\,,
\end{eqnarray}
\begin{equation}\label{KGequation}
2\,\omega(\phi)\,\Box\phi + \omega_\phi(\phi)\,\phi_{;\alpha}\phi^{;\alpha} - f_\phi = 0\,,
\end{equation}
where the shorthand notation $f_R = \partial f/\partial R$, $f_Y = \partial f/\partial Y$, $f_\phi = \partial f/\partial \phi$, $\omega_\phi = d\omega/d\phi$ is adopted, $\Box = \nabla_\mu \nabla^\mu$ is the d'Alembert operator, and 
\begin{equation}
T_{\mu\nu} = -\,\frac{2}{\sqrt{-g}}\,\frac{\delta\bigl(\sqrt{-g}\,\mathcal{L}_m\bigr)}{\delta g^{\mu\nu}}
\end{equation}
is the energy-momentum tensor of matter. Taking the trace of Eq.~\eqref{fieldequationSTFOG} one obtains the master equation
\begin{equation}\label{traceSTFOG}
f_R\,R + 2\,f_Y\,R_{\alpha\beta}R^{\alpha\beta} - 2\,f + \Box\bigl[3\,f_R + f_Y\,R\bigr] + 2\bigl[f_Y\,R^{\alpha\beta}\bigr]_{;\alpha\beta} - \omega(\phi)\,\phi_{;\alpha}\phi^{;\alpha} = \mathcal{X}\,T\,,
\end{equation}
with $T = T^{\sigma}{}_{\sigma}$.

%------------------------------------------------------
\subsection{NCSG as a Weyl-squared specialization of STFOG}\label{sec:NCSG}

Non-Commutative Spectral Geometry (NCSG) constitutes a special case of STFOG that has attracted considerable attention as a candidate framework for unifying all fundamental interactions, including gravity, within a purely geometric setting \cite{connes_1,ccm,mairi2012}. In the NCSG approach, the geometry is described by the tensor product $\mathcal{M}\times\mathcal{F}$ of a four-dimensional compact Riemannian manifold $\mathcal{M}$ and a discrete non-commutative space $\mathcal{F}$ encoding the internal degrees of freedom of the Standard Model of particle physics \cite{Chamseddine:2005zk,Chamseddine:2008zj,cchiggs,Chamseddine:2013rta}.

At the grand unification scale (set by the cutoff $\Lambda$), the gravitational part of the bosonic sector of the NCSG spectral action, in Lorentzian signature, reads \cite{Nelson:2008uy,mairi2012}
\begin{equation}\label{NCSG-action}
\mathcal{S}_{\text{grav}}^{\text{NCSG}} = \int d^4x\,\sqrt{-g}\,\biggl[\frac{R}{2\mathcal{X}} + \alpha_0\,C_{\mu\nu\rho\sigma}C^{\mu\nu\rho\sigma} + \tau_0\,R^\star R^\star - \xi_0\,R\,|\mathbf{H}|^2\biggr]\,,
\end{equation}
where $\mathcal{X} = 8\pi G/c^4$, $\mathbf{H} = (\sqrt{a\,f_0}/\pi)\,\phi$ is the Higgs field, $C_{\mu\nu\rho\sigma}$ is the Weyl tensor, $R^\star R^\star$ is the topological Gauss--Bonnet term, $\alpha_0 = -3f_0/(10\pi^2)$, and $\xi_0 = 1/12$. The term $R^\star R^\star$ does not contribute to the field equations, while the square of the Weyl tensor can be decomposed as
\begin{equation}\label{Weyl-decomposition}
C_{\mu\nu\rho\sigma}\,C^{\mu\nu\rho\sigma} = 2\,R_{\mu\nu}R^{\mu\nu} - \tfrac{2}{3}\,R^2\,.
\end{equation}
Equation~\eqref{Weyl-decomposition} holds up to the topological Gauss--Bonnet
combination of Eq.~\eqref{GaussBonnet}: the exact four-dimensional identity is
$C_{\mu\nu\rho\sigma}C^{\mu\nu\rho\sigma}
 = \mathcal{G} + 2\,R_{\mu\nu}R^{\mu\nu} - \tfrac{2}{3}\,R^2$,
and since $\mathcal{G}$ does not contribute to the field equations it has been
discarded.
The action \eqref{NCSG-action} is embedded in the STFOG class \eqref{FOGaction} as a constrained Weyl-squared specialization rather than as a generic regular point of the full scalar-tensor parameter space. Neglecting the non-minimal coupling $\xi_0\,R\,|\mathbf{H}|^2$ (an approximation which preserves the homogeneous and isotropic cosmological sector \cite{Lambiase:2013dai}), the variation with respect to $g_{\mu\nu}$ yields the field equation
\begin{equation}\label{NCSG-FE}
G_{\mu\nu} + \frac{1}{\beta^2}\Bigl[2\,\nabla_\lambda\nabla_\kappa C_{\mu}{}^{\lambda}{}_{\nu}{}^{\kappa} + C_{\mu}{}^{\lambda}{}_{\nu}{}^{\kappa}R_{\lambda\kappa}\Bigr] = \mathcal{X}\,T_{\mu\nu}\,,
\end{equation}
where $G_{\mu\nu}$ is the Einstein tensor and $\beta^2 = 5\pi^2/(6\,\mathcal{X}\,f_0)$. The parameter $f_0$ is the zeroth moment of the cutoff function of the spectral action; current experimental bounds from Gravity Probe B and torsion-balance experiments yield $f_0 \gtrsim 10^{24}$ \cite{Lambiase:2013dai,eot}, while analysis of binary pulsars provides complementary constraints \cite{NaefJetzer2010}. The effective mass associated with the NCSG correction is $\beta = \sqrt{5\pi^2/(6\,\mathcal{X}\,f_0)}$, which governs the range of the Yukawa-type modification to the gravitational potentials, as we shall show in Sec.~\ref{sec:linearized_field_equations}.

At the action level, Eqs.~\eqref{NCSG-action}--\eqref{Weyl-decomposition} identify NCSG as a special constrained sector of STFOG. A structural property of the field equation~\eqref{NCSG-FE} that will be important in what follows is that its bracket is twice the Bach tensor $B_{\mu\nu}$, which is symmetric, divergence-free and identically traceless, i.e. $B^{\mu}{}_{\mu}\equiv 0$. Taking the trace of Eq.~\eqref{NCSG-FE} therefore yields the GR-like relation $R = -\mathcal{X}\,T$, so that NCSG propagates no scalar mode (scalaron) and $R=0$ in vacuum.

\subsection{Weak field expansion and the Standard Post-Newtonian gauge}\label{sec:PNformalism}
\textit{1) Expansion of the metric.}\;
We work in the weak-field, slow-motion regime and expand the metric tensor about the Minkowski background $\eta_{\mu\nu} = \text{diag}(+1,-1,-1,-1)$ as in \cite{Stabile2014,FOG1,FOG2,Will}. Let us introduce the dimensionless expansion parameter
$\varepsilon \equiv \bar v/c \ll 1$, where $\bar v$ is a characteristic orbital velocity; for virialised systems the virial estimate $\bar v^{2}\sim|\Phi|$ gives $|\Phi|/c^2 = \mathcal{O}(\varepsilon^2)$, and the weak-field and slow-motion expansions are consistently organized in powers of $\varepsilon$.

The metric expansion reads
\begin{equation}\label{PN-metric}
\begin{aligned}
g_{00} &\sim 1 + {}^{(2)}\!g_{00} + {}^{(4)}\!g_{00} + O(\varepsilon^{6})\,,\\[3pt]
g_{0i} &\sim {}^{(3)}\!g_{0i} + O(\varepsilon^{5})\,,\\[3pt]
g_{ij} &\sim -\delta_{ij} + {}^{(2)}\!g_{ij} + O(\varepsilon^{4})\,,
\end{aligned}
\end{equation}
where the left superscript in parentheses, ${}^{(n)}$, denotes a quantity of order $O(\varepsilon^n)$ on every tensor and scalar entering the post-Newtonian expansion (equivalently, $n$-th order in $c^{-1}$); the same convention applies to the scalar field, which is expanded as $\phi \sim \phi^{(0)} + \varphi + \ldots$, with $\varphi = O(\varepsilon^2)$, and will be used below for curvature terms ${}^{(n)}\!R_{\mu\nu}$, ${}^{(n)}\!R$ and for the energy-momentum components ${}^{(n)}\!T_{\mu\nu}$.
Taking into account that spatial and time derivatives scale as $\partial/\partial x^i \sim 1/\bar{r}$ and $\partial/\partial x^0 \sim \bar{v}/\bar{r} \sim \varepsilon/\bar{r}$, we identify the gravitational potentials through
\begin{equation}\label{potentials-def}
{}^{(2)}\!g_{00} = \frac{2\,\Phi}{c^2}\,,\qquad {}^{(2)}\!g_{ij} = \frac{2}{c^2}\bigl(\Psi\,\delta_{ij} + \chi_{,ij}\bigr)\,,\qquad {}^{(3)}\!g_{0i} = \frac{Z_i}{c^3}\,.
\end{equation}
If one considers the weak-field limit, ${}^{(4)}\!g_{00}$ and the longitudinal part $\chi_{,ij}$ can be neglected in the metric, so that the linearized line element takes the isotropic form\footnote{The scalar $\chi$ is the longitudinal part of the spatial metric. This is fixed by the gauge condition \eqref{SPN-gauge-explicit} to satisfy $\nabla^2\chi = \Psi - \Phi$, and it vanishes identically in the GR-like regime $\gamma=1$ ($\Psi=\Phi$). More generally, $\chi_{,ij}=O(\Psi-\Phi)$, and the difference $\Psi-\Phi$ contains only short-ranged Yukawa terms, which are exponentially suppressed at the scales of interest in the screened regime $\Phi\sim\Psi$ ($\gamma\sim1$) of the viable theories considered here (see Sec.~\ref{sec:brumberg} and Ref.~\cite{TedescoCapolupoLambiase_EPJC2024}); this contribution is therefore negligible at the retained order. Thus, the isotropic line element \eqref{ds2-PN} remains valid although $\Phi$ and $\Psi$ are not analytically identical.
}
\begin{equation}\label{ds2-PN}
ds^2 = \biggl(1 + \frac{2\Phi}{c^2}\biggr)\,c^2\,dt^2 + \frac{2\,Z_i}{c^3}\,c\,dt\,dx^i - \biggl(1 - \frac{2\Psi}{c^2}\biggr)\,\delta_{ij}\,dx^i\,dx^j + O(\varepsilon^{4})\,.
\end{equation}
The gravitomagnetic potential $Z_i$ admits the decomposition $Z_i = A_i - \partial^2 X/(\partial t\,\partial x^i)$, where $A_i$ is the vector potential and $X$ is the superpotential satisfying $\nabla^2 X = \Phi$. 
\medskip

\textit{2) Standard Post-Newtonian gauge.}\; For our purposes, the appropriate coordinate choice is the \textit{Standard Post-Newtonian (SPN) gauge}\footnote{This gauge is especially suitable for deriving the $N$-body equations of motion, as it directly provides the potentials entering the Lagrangian \cite{Brumberg,BrumbergBook1991,Brumberg2,Kopeikin2011,Soffel}.}, defined by the conditions \cite{Will,Brumberg}
\begin{equation}\label{SPN-gauge}
g_{0j,j} - \tfrac{1}{2}\,g_{jj,0} = O(\varepsilon^{5})\,,\qquad\quad
g_{ij,j} - \tfrac{1}{2}\,(g_{jj} - g_{00})_{,i} = O(\varepsilon^{4})\,,
\end{equation}
which, at the relevant orders, implies
\begin{equation}\label{SPN-gauge-explicit}
{}^{(3)}\!g_{0h,h} - \tfrac{1}{2}\,{}^{(2)}\!g_{hh,0} = 0\,,\qquad\quad
{}^{(2)}\!g_{ij,j} + \tfrac{1}{2}\,{}^{(2)}\!g_{00,i} - \tfrac{1}{2}\,{}^{(2)}\!g_{jj,i} = 0\,.
\end{equation}
By differentiation of these conditions one derives the auxiliary relations
\begin{equation}\label{SPN-derived}
{}^{(3)}\!g_{0h,hi} - \tfrac{1}{2}\,{}^{(2)}\!g_{hh,0i} = 0\,,\qquad
{}^{(3)}\!g_{0h,0h} - \tfrac{1}{2}\,{}^{(2)}\!g_{hh,00} = 0\,,\qquad
{}^{(2)}\!g_{00,ih} + {}^{(2)}\!g_{ij,jh} + {}^{(2)}\!g_{hj,ji} - {}^{(2)}\!g_{jj,ih} = 0\,.
\end{equation}
One may alternatively adopt the harmonic (de Donder) gauge, characterized by $g^{\mu\nu}\Gamma^\alpha_{\mu\nu} = 0$. The transformation connecting the two gauges is $x_0 = \tilde{x}_0 + X_{,0}$, $x^i = \tilde{x}^i$, under which $g_{00} \to g_{00} + 2\,X_{,00}$ and $g_{0i} \to g_{0i} + X_{,0i}$, while $g_{ij}$ is unchanged. This yields, in harmonic coordinates, $g_{0i} = A_i$, so that the cross-term $Z_i$ reduces to the vector potential alone.
\medskip

\textit{3) Ricci tensor and Ricci scalar in SPN gauge.}\; After applying the SPN gauge conditions \eqref{SPN-gauge-explicit}--\eqref{SPN-derived} to the general post-Newtonian expressions, the components of the Ricci tensor simplify to\footnote{Here and in the following, repeated Latin indices are summed over $1,2,3$, $\nabla^2 = \partial_{hh}$ denotes the flat-space Laplacian.}
\begin{equation}\label{Ricci-SPN}
\begin{aligned}
{}^{(2)}\!R_{00} &= \tfrac{1}{2}\,\nabla^2{}^{(2)}\!g_{00}\,,\\[4pt]
{}^{(4)}\!R_{00} &= \tfrac{1}{2}\,\nabla^2{}^{(4)}\!g_{00} + \tfrac{1}{2}\,{}^{(2)}\!g_{hl}\,{}^{(2)}\!g_{00,hl} + \tfrac{1}{2}\,{}^{(2)}\!g_{hl,h}\,{}^{(2)}\!g_{00,l} - \tfrac{1}{2}\,|\nabla{}^{(2)}\!g_{00}|^2\,,\\[4pt]
{}^{(3)}\!R_{0i} &= \frac{1}{2}\left(\nabla^2{}^{(3)}\!g_{0i} - {}^{(3)}\!g_{0h,hi} - {}^{(2)}\!g_{ih,0h} + {}^{(2)}\!g_{hh,0i} \right)\,,\\[4pt]
{}^{(2)}\!R_{ij} &= \tfrac{1}{2}\,\nabla^2{}^{(2)}\!g_{ij}\,.
\end{aligned}
\end{equation}
The corresponding Ricci scalar at second and fourth order reads
\begin{equation}\label{Ricci-scalar-SPN}
\begin{aligned}
{}^{(2)}\!R &= \tfrac{1}{2}\,\nabla^2{}^{(2)}\!g_{00} - \tfrac{1}{2}\,\nabla^2{}^{(2)}\!g_{hh}\,,\\[4pt]
{}^{(4)}\!R &= \tfrac{1}{2}\,\nabla^2{}^{(4)}\!g_{00} + \tfrac{1}{2}\,{}^{(2)}\!g_{hl}\,{}^{(2)}\!g_{00,hl} + \tfrac{1}{2}\,{}^{(2)}\!g_{hl,h}\,{}^{(2)}\!g_{00,l} - \tfrac{1}{2}\,|\nabla{}^{(2)}\!g_{00}|^2
- \tfrac{1}{2}\,{}^{(2)}\!g_{00}\,\nabla^2{}^{(2)}\!g_{00} - \tfrac{1}{2}\,{}^{(2)}\!g_{hl}\,\nabla^2{}^{(2)}\!g_{hl}\,.
\end{aligned}
\end{equation}
This is the SPN-gauge expression used below in the scalar sector and in the NCSG specialization.

Using the definitions of the post-Newtonian potentials \eqref{potentials-def} together with the SPN gauge relations \eqref{SPN-gauge-explicit}, the Ricci tensor components \eqref{Ricci-SPN} take the form
\begin{equation}\label{ricci-potentials}
{}^{(2)}\!R_{00} = \tfrac{1}{c^2}\, \nabla^2\Phi\,,\qquad
{}^{(3)}\!R_{0i} = \tfrac{1}{2c^3}\bigl(\nabla^2 Z_i + \Psi_{,0i}\bigr)\,,\qquad
{}^{(2)}\!R_{ij} = \tfrac{1}{c^2}\nabla^2\Psi\,\delta_{ij} + \tfrac{1}{c^2}(\Psi - \Phi)_{,ij}\,\,\,\,\,\,,
\end{equation}
and the second-order Ricci scalar is
\begin{equation}\label{Ricci-scalar-SPN-2}
{}^{(2)}\!R
=
\frac{1}{c^2}
\left(
2\nabla^2\Phi-4\nabla^2\Psi
\right).
\end{equation}

\medskip
\textit{4) Energy-momentum tensor for point particles.}\; In order to solve the field equations, for a system of $N$ gravitating point particles with masses $m_a$ ($a = 1,\ldots,N$), neglecting pressure and internal energy\footnote{This approximation is appropriate for celestial bodies at distances much larger than their own radii, as in the Solar System, binary pulsars, or stellar clusters.}, the energy-momentum tensor of a pressureless perfect fluid, $T_{\mu\nu} = \rho\,c^2\,u_\mu\,u_\nu$ with $u^\sigma u_\sigma = 1$, has the contravariant form
\begin{equation}\label{Tmunu-contra}
T^{\mu\nu}(\mathbf{x},t) =
\frac{1}{\sqrt{-g}}\,
\sum_{a=1}^{N}\gamma_a\,m_a\,
\frac{dx_a^\mu}{dt}\,
\frac{dx_a^\nu}{dt}\,
\delta(\mathbf{x} - \mathbf{x}_a(t))\,,
\end{equation}
with $g=\text{det}\, g_{\mu\nu}$, $m_a$ the mass of the $a$-th particle, $\tau_a$ its proper time, $\mathbf{x}_a(t)$ its spatial position at coordinate time $t$, and $\gamma_a = (d\tau_a/dt)^{-1}$. This distributional source is supported on the timelike worldlines $x_a^{\mu}(\tau_a)$.
Since the linearized field equations in the SPN gauge involve the components $T_{\mu\nu}$ with both indices lowered, one needs the covariant form. Therefore, expanding to the required post-Newtonian accuracy, as a first step one finds the contravariant components \cite{Weinberg}. Then, using $T_{\mu\nu} = g_{\mu\alpha}\,g_{\nu\beta}\,T^{\alpha\beta}$ and the leading-order relations $g_{00} \sim 1$, $g_{ij} \sim -\delta_{ij}$, one obtains at the required post-Newtonian order
\begin{equation}\label{Tmunu-covariant}
{}^{(0)}\!T_{00} = \sum_a m_a\,c^2\,\delta(\mathbf{x} - \mathbf{x}_a)\,,\qquad
{}^{(1)}\!T_{0i} = -\sum_a m_a\,c\,v^{a}_{i}\,\delta(\mathbf{x} - \mathbf{x}_a)\,, \qquad
{}^{(0)}\!T_{ij} = 0\,,
\end{equation}
while the trace is $T = T^{\sigma}{}_{\sigma} = g^{\mu\nu}T_{\mu\nu} \sim \sum_a m_a\,c^2\,\delta(\mathbf{x} - \mathbf{x}_a)$ at leading order. Note the sign change in $T_{0i}$ with respect to $T^{0i}$, which arises from $g_{ij} \sim -\delta_{ij}$ through the index-lowering operation $T_{0i} = g_{00}\,g_{ij}\,T^{0j} \sim (-\delta_{ij})\,T^{0j} = -T^{0i}$. The expressions \eqref{Tmunu-covariant} constitute the point-like sources for the STFOG system.

In Sec.~\ref{sec:brumberg}, we show that the linearized metric in Eq.~\eqref{ds2-PN}, i.e. the weak-field limit, is sufficient for the present construction. In the physical regimes that ensure $\Phi\sim\Psi$ for the class of theories considered here, the fourth-order term ${}^{(4)}\!g_{00}$ turns out to be dynamically irrelevant for the ETG $N$-body system, as also happens in General Relativity. 

%------------------------------------------------------
%------------------------------------------------------
\section{PPN Constraints for Extended Gravity}\label{sec:brumberg}

The standard post-Newtonian expansion \eqref{PN-metric} involves, for the $00$-component, both the second-order perturbation ${}^{(2)}\!g_{00}$ and the fourth-order one ${}^{(4)}\!g_{00}$. Whether the latter is actually required for deriving the first post-Newtonian equations of motion is a central question, particularly in extended theories of gravity where the proliferation of higher-order terms makes the calculation of ${}^{(4)}\!g_{00}$ considerably more involved than in GR. In this section we show that, under the observationally well-supported hypothesis $\gamma \sim 1$, the linearized metric components ${}^{(2)}\!g_{00},\,{}^{(3)}\!g_{0i}$, and ${}^{(2)}\!g_{ij}$ are sufficient at 1PN order for the full STFOG class.

%------------------------------------------------------
\subsection{PPN bounds, screening, and the \texorpdfstring{$\gamma \sim 1$}{gamma ~ 1} regime}\label{sec:Eddington-Robertson}

For a static, spherically symmetric source of mass $M$, the linearized metric outside the source can be written, without assuming specific field equations, as the Eddington--Robertson expansion \cite{Eddington,Eddington2,Robertson,Gravitation,Will,Weinberg}
\begin{equation}\label{Robertson-expansion}
g_{00} = 1 - \frac{2GM}{c^2\,r} + 2\,(\beta_{\text{\tiny PPN}} - \gamma)\,\biggl(\frac{GM}{c^2\,r}\biggr)^{\!2} + O(\varepsilon^{6})\,,
\end{equation}
\begin{equation}\label{Robertson-expansion-ij}
g_{ij} = -\biggl(1 + 2\,\gamma\,\frac{GM}{c^2\,r}\biggr)\,\delta_{ij} + O(\varepsilon^{4})\,.
\end{equation}
The two dimensionless parameters $\gamma$ and $\beta_{\text{\tiny PPN}}$ encode, respectively, the amount of spatial curvature produced per unit rest mass and the degree of nonlinearity in the gravitational superposition law \cite{Will,Soffel,Gravitation}\footnote{Here and in the following we use $\beta_{\text{\tiny PPN}}$ for the second PPN parameter, to avoid confusion with the NCSG mass parameter $\beta$ introduced in Sec.~\ref{sec:NCSG}.}. In terms of the potentials defined in Eq.~\eqref{potentials-def}, the first PPN parameter reads $\gamma = \Psi/\Phi$, so that $\gamma = 1$ corresponds to $\Psi = \Phi$, as in GR. Consistency with the Schwarzschild solution requires $\gamma = \beta_{\text{\tiny PPN}} = 1$. As noted by Weinberg \cite{Weinberg}, once $\gamma \sim 1$ is established from light-propagation experiments, the observed perihelion precession of Mercury constrains
\begin{equation}\label{beta-gamma-relation}
\beta_{\text{\tiny PPN}} \sim 2\gamma - 1\,.
\end{equation}
The PPN formalism was subsequently extended by Nordtvedt, Will, Thorne, and Misner \cite{Nordtvedt,Will,Soffel,Gravitation} into a comprehensive framework for testing any metric theory of gravity against Solar-System observations \cite{Shapiro,Nordtvedt}. The current experimental bounds, summarized in Table~\ref{tab:PPN-bounds} and comprehensively reviewed in Ref.~\cite{Will2014}, demonstrate that both parameters are consistent with GR at the level of a few parts in $10^5$.

\renewcommand{\arraystretch}{2.0}
\begin{table}[ht]
\centering
\begin{tabular}{@{\hspace{0.2cm}}l@{\hspace{0.6cm}}c@{\hspace{0.6cm}}l@{\hspace{0.6cm}}l@{\hspace{0.2cm}}}
\specialrule{1.5pt}{0pt}{0pt}
\textbf{PPN parameter} & \textbf{Bound} & \textbf{Physical effect} & \textbf{Experiment} \\
\midrule
$|\gamma - 1|$ & $(2.1 \pm 2.3) \times 10^{-5}$ & Time delay, light deflection & Cassini tracking \\
$|\beta_{\text{\tiny PPN}} - 1|$ & $(1.6 \pm 1.8) \times 10^{-5}$ & Nordtvedt effect, perihelion & MESSENGER \\
$|\beta_{\text{\tiny PPN}} - 1|$ & $(1.2 \pm 1.1) \times 10^{-4}$ & Nordtvedt effect & Lunar laser ranging \\
\specialrule{1.5pt}{0pt}{0pt}
\end{tabular}
\caption{Current bounds on the PPN parameters from Solar-System experiments. The tightest constraint on $\gamma$ comes from the Cassini spacecraft Shapiro time-delay measurement \cite{Cassini}; the most precise bound on $\beta_{\text{\tiny PPN}}$ is obtained from MESSENGER radio tracking of Mercury \cite{Genova2018}, while the Lunar Laser Ranging bound is reported in Ref.~\cite{Williams2004}.}
\label{tab:PPN-bounds}
\end{table}

In the context of ETG, the PPN parameters are generally different from their GR values. In particular, $f(R)$-gravity, being dynamically equivalent to a Brans--Dicke theory with $\omega_{\text{BD}} = 0$ \cite{CapozzielloFaraoni,DeFelice:2010aj}, yields $\gamma = 1/2$ in the regime $m_R\,r \ll 1$, where $m_R$ is the effective mass of the scalar degree of freedom (the scalaron). 

Such a value would immediately rule out the theory (see Table~\ref{tab:PPN-bounds}). However, all curvature-based ETG possessing a scalar degree of freedom exhibit \textit{chameleon screening} \cite{71,72,73,73a,74}: the effective mass of the scalaron depends on the local matter density and curvature through the effective potential $V_{\text{eff}}(\phi) = V(\phi) + \rho\,\ln A(\phi)$, so that $m_{\text{eff}}^2 = V''_{\text{eff}}(\phi_{\min})$ is an increasing function of the ambient density $\rho$. In high-density environments such as the Solar System, the scalaron mass becomes large ($m_R\,r \gg 1$), and the Yukawa correction $\propto e^{-m_R r}/r$ is exponentially suppressed at the experimentally accessible scales, restoring $\gamma \sim 1$. Conversely, at cosmological distances the scalar field becomes long-range and can drive the observed accelerated expansion \cite{HuSawicki,DeFelice:2010aj}. Quantitatively, the STFOG gravitational potential for a point source takes the general form~\cite{TedescoCapolupoLambiase_EPJC2024}
\begin{equation}\label{Phi-Yukawa}
\Phi(r) = -\frac{GM}{r}\biggl[1 + \sum_{i = \pm,\,Y} F_i\,e^{-\beta_i\,r}\biggr]\,,
\end{equation}
where $F_i$ are dimensionless coupling strengths and $\beta_i$ are the inverse range parameters (effective masses) of the Yukawa modes.

The observational bounds on these parameters, inferred from the periastron advance of Solar-System planets and the S2 star around Sagittarius~A$^*$ \cite{TedescoCapolupoLambiase_EPJC2024}, are collected in Table~\ref{tab:bounds-STFOG-NCSG}. As shown in Ref.~\cite{TedescoCapolupoLambiase_EPJC2024}, for representative values $|F_i| \sim 10^{-1}$ and $\beta_i\,r \gtrsim 10$ the percentage deviation between $\Phi$ and $\Psi$ is $\sim 10^{-5}\,\%$, rendering $\gamma$ virtually indistinguishable from unity within the current experimental precision. These bounds show that, in the systems of interest, all Yukawa corrections are either short-ranged ($\beta_i\,r \gg 1$) or weakly coupled ($|F_i| \ll 1$); we refer to this situation as the \textit{screened regime}. In the scalar sector the suppression is enforced dynamically by the chameleon mechanism, while for the Ricci-tensor mode it is imposed observationally; in either case $\gamma \sim 1$ and $\beta_{\text{\tiny PPN}} \sim 1$ follow in the regimes relevant to post-Newtonian dynamics.

\renewcommand{\arraystretch}{2.00}
\begin{table}[ht]
\centering
\setlength{\tabcolsep}{10pt}
\begin{tabular}{@{}lccc@{}}
\specialrule{1.5pt}{0pt}{0pt}
\textbf{Object} 
& $|\eta|$ 
& \textbf{STFOG Constraint} 
& \textbf{NCSG Constraint} \\
\midrule
Mercury 
& $0.5\,^{\prime\prime}\,\mathrm{cy}^{-1}$ 
& $|F_i| \lesssim 3.44\times10^{-12}$ 
& $\beta \gtrsim 5.15\times10^{-10}\,\mathrm{m}^{-1}$ \\

Mars 
& $5\times10^{-4}\,^{\prime\prime}\,\mathrm{cy}^{-1}$ 
& $|F_i| \lesssim 2.69\times10^{-11}$ 
& $\beta \gtrsim 1.46\times10^{-10}\,\mathrm{m}^{-1}$ \\

Jupiter 
& $4\times10^{-3}\,^{\prime\prime}\,\mathrm{cy}^{-1}$ 
& $|F_i| \lesssim 1.36\times10^{-9}$ 
& $\beta \gtrsim 3.61\times10^{-11}\,\mathrm{m}^{-1}$ \\

S2\,(Sgr\,A$^*$) 
& $0.014\,^\circ$ 
& $|F_i| \lesssim 5.15\times10^{-4}$ 
& $\beta \gtrsim 5.43\times10^{-13}\,\mathrm{m}^{-1}$ \\
\specialrule{1.5pt}{0pt}{0pt}
\end{tabular}
\caption{Observational constraints on the STFOG coupling strengths $|F_i|$ and on the NCSG mass parameter $\beta$, derived from the relativistic periastron advance of Solar-System planets and from the S2 star orbit around Sagittarius~A$^*$ \cite{TedescoCapolupoLambiase_EPJC2024}. Here $|\eta|$ denotes the observational upper bound on the periastron advance used to constrain deviations from the GR prediction. The values of $|\eta|$ include the appropriate units, namely $^{\prime\prime}\,\mathrm{cy}^{-1}$ for Solar-System planets and $^\circ$ for S2.}
\label{tab:bounds-STFOG-NCSG}
\end{table}

%------------------------------------------------------
\subsection{Brumberg's conjecture beyond General Relativity}\label{sec:Brumberg-conjecture}

\subsubsection{Brumberg cancellation in General Relativity}

In GR, the post-Newtonian equations of motion of $N$ gravitating bodies, i.e. the celebrated Einstein--Infeld--Hoffmann (EIH) equations \cite{EIH} --- heir to the early relativistic $N$-body analyses of J. Droste \cite{Droste} and T. Levi-Civita \cite{LeviCivita} --- are customarily derived from the geodesic equation for the particle worldlines, which requires the knowledge of ${}^{(4)}\!g_{00}$ through the Christoffel symbols at 1PN order. However, as originally pointed out by Infeld \cite{Infeld,Infeld2} and systematically developed by Brumberg \cite{Brumberg,BrumbergBook1991,Brumberg2,Brumberg3}, the equations of motion can alternatively be derived from the \textit{variational principle} for the gravitational field equations. With this approach it was demonstrated that the EIH Lagrangian and the resulting equations of motion are entirely independent of ${}^{(4)}\!g_{00}$. This is the content of the \textit{Brumberg conjecture}: the linearized metric \eqref{ds2-PN} is sufficient for relativistic celestial mechanics and astrometry within GR.

\subsubsection{Post-Newtonian counting and the suppression of \texorpdfstring{${}^{(4)}\!g_{00}$}{(4)g00}}

In ETG, the field equations are of fourth order and involve additional degrees of freedom, so the extension of the Brumberg result requires a dedicated analysis. We now demonstrate, on the basis of the Eddington--Robertson expansion and the experimentally established values of the PPN parameters, that ${}^{(4)}\!g_{00}$ is physically irrelevant at 1PN order in the STFOG class under the observationally supported condition $|\gamma - 1| \ll 1$.

Comparing the Robertson expansion \eqref{Robertson-expansion}--\eqref{Robertson-expansion-ij} with the post-Newtonian metric \eqref{PN-metric}, one identifies
\begin{equation}\label{metric-components-identification}
{}^{(2)}\!g_{00} = -\frac{2\,GM}{c^2\,r}\,,\qquad
{}^{(2)}\!g_{ij} = - \gamma\,\frac{2\,GM}{c^2\,r}\,\delta_{ij}\,,\qquad
{}^{(4)}\!g_{00} = 2\,(\beta_{\text{\tiny PPN}} - \gamma)\,\biggl(\frac{GM}{c^2\,r}\biggr)^{\!2}\,.
\end{equation}
Inserting the relation \eqref{beta-gamma-relation} and defining the deviation from the GR value by $\delta\gamma \equiv \gamma - 1$, with $|\delta\gamma| \lesssim 2.3 \times 10^{-5}$ (Cassini bound), one obtains $\beta_{\text{\tiny PPN}} - \gamma \sim \delta\gamma$, so that
\begin{equation}\label{4g00-estimate}
{}^{(4)}\!g_{00} \sim 2\,\delta\gamma\,\biggl(\frac{GM}{c^2\,r}\biggr)^{\!2}\,.
\end{equation}
For gravitationally bound systems the virial theorem gives $\varepsilon^2 \equiv v^2/c^2 \sim GM/(c^2\,r)$, where $\varepsilon = v/c$ is the velocity parameter. The resulting post-Newtonian hierarchy of the metric perturbations then reads
\begin{equation}\label{PN-hierarchy}
|{}^{(2)}\!g_{00}|\sim|{}^{(2)}\!g_{ij}|\sim\varepsilon^{2}\;\gg\;|{}^{(3)}\!g_{0i}|\sim\varepsilon^{3}\;\gg\;|{}^{(4)}\!g_{00}|\sim|\delta\gamma|\,\varepsilon^{4}\,,
\end{equation}
and, besides the usual post-Newtonian suppression, the further $|\delta\gamma|$ factor reduces ${}^{(4)}\!g_{00}$ by several orders of magnitude whenever $|\delta\gamma|\ll 1$. In the same small-deviation regime, it is worth stressing that the linearized spatial components ${}^{(2)}\!g_{ij}$, which enter the 1PN dynamics on equal footing with ${}^{(2)}\!g_{00}$, are themselves much larger than ${}^{(4)}\!g_{00}$.

Table~\ref{tab:PN-hierarchy} collects the numerical magnitudes for a few representative astrophysical systems. Already for Mercury, $|{}^{(4)}\!g_{00}|$ is about $10^{-20}$, to be compared with $|{}^{(2)}\!g_{00}|\sim|{}^{(2)}\!g_{ij}|\sim 5\times 10^{-8}$ and $|{}^{(3)}\!g_{0i}|\sim 4\times 10^{-12}$, so it is suppressed by twelve and eight orders of magnitude respectively. Even for the S2 star, which is the most relativistic system considered here ($\varepsilon\sim 0.02$), $|{}^{(4)}\!g_{00}|/|{}^{(2)}\!g_{00}|$ does not exceed $\sim 10^{-8}$.

\renewcommand{\arraystretch}{2.0}
\begin{table}[ht]
\centering
\begin{tabular}{@{\hspace{0.15cm}}l@{\hspace{0.4cm}}c@{\hspace{0.4cm}}c@{\hspace{0.4cm}}c@{\hspace{0.4cm}}c@{\hspace{0.4cm}}c@{\hspace{0.15cm}}}
\specialrule{1.5pt}{0pt}{0pt}
\textbf{Object} & $\varepsilon = v/c$ & $|{}^{(2)}\!g_{00}| \approx |{}^{(2)}\!g_{ij}|$ & $|{}^{(3)}\!g_{0i}|$ & $|{}^{(4)}\!g_{00}|_{\mathrm{Cassini}}$ & $|{}^{(4)}\!g_{00}|/|{}^{(2)}\!g_{00}|$ \\
\midrule
Mercury & $1.6\!\times\!10^{-4}$ & $5\!\times\!10^{-8}$ & $4\!\times\!10^{-12}$ & $3\!\times\!10^{-20}$ & $6\!\times\!10^{-13}$ \\
Earth & $1.0\!\times\!10^{-4}$ & $2\!\times\!10^{-8}$ & $10^{-12}$ & $4\!\times\!10^{-21}$ & $2\!\times\!10^{-13}$ \\
Saturn & $3.2\!\times\!10^{-5}$ & $2\!\times\!10^{-9}$ & $3\!\times\!10^{-14}$ & $5\!\times\!10^{-23}$ & $2\!\times\!10^{-14}$ \\
S2\,(Sgr\,A$^*$) & $2\!\times\!10^{-2}$ & $8\!\times\!10^{-4}$ & $8\!\times\!10^{-6}$ & $7\!\times\!10^{-12}$ & $9\!\times\!10^{-9}$ \\
\specialrule{1.5pt}{0pt}{0pt}
\end{tabular}
\caption{Post-Newtonian hierarchy of the metric perturbations for representative astrophysical systems, adopting the Cassini bound $|\delta\gamma| \lesssim 2.3\times 10^{-5}$. The magnitude of ${}^{(4)}\!g_{00}$ is evaluated from Eq.~\eqref{4g00-estimate}. In all cases $|{}^{(4)}\!g_{00}|$ is negligible compared to all linearized components, including ${}^{(2)}\!g_{ij}$.}
\label{tab:PN-hierarchy}
\end{table}

The irrelevance of ${}^{(4)}\!g_{00}$ can be made more precise at the level of the equations of motion. The contribution of ${}^{(4)}\!g_{00}$ to the acceleration is
\begin{equation}\label{accel-4g00}
\delta a^i\bigl|_{{}^{(4)}\!g_{00}} \sim \frac{c^2}{2}\,\partial_i\,{}^{(4)}\!g_{00} \sim \frac{(\delta\gamma)\,(GM)^2}{c^2\,r^3}\,,
\end{equation}
while the Newtonian (leading) acceleration is $a^i_{\text{N}} = GM/r^2$. Their ratio gives
\begin{equation}\label{ratio-4g00}
\frac{|\delta a^i|_{{}^{(4)}\!g_{00}}}{|a^i_{\text{N}}|} \sim |\delta\gamma|\,\varepsilon^2\,.
\end{equation}
For Mercury, this ratio is $\lesssim 6\times 10^{-13}$; for the S2 star, $\lesssim 9\times 10^{-9}$. Both values lie well below the present observational accuracy; they are equally small if compared with the 1PN correction itself, which is of order $\varepsilon^2$ relative to the Newtonian acceleration, since $|\delta a^i|_{{}^{(4)}\!g_{00}}/|a^i_{\text{\tiny 1PN}}| \sim |\delta\gamma| \lesssim 10^{-5}$.

Under the hypothesis $\Phi \sim \Psi$, that is $\delta\gamma = \gamma - 1$ with $|\delta\gamma| \ll 1$ (a physical requirement characterizing the screened regime of viable ETG, see Sec.~\ref{sec:Eddington-Robertson} and Table~\ref{tab:bounds-STFOG-NCSG}), the linearized metric \eqref{ds2-PN} captures all the content of the 1PN equations of motion in the full STFOG class. An independent confirmation will follow in Sec.~\ref{sec:relativistic_dynamics} from the variational construction of the $N$-body Lagrangian, where the cancellation of ${}^{(4)}\!g_{00}$ is displayed explicitly. The validity range of the present framework extends to the Solar System, hierarchical stellar systems, binary pulsars, and stellar orbits around supermassive black holes \cite{TedescoCapolupoLambiase_EPJC2024,GRAVITY2020}. Systems that need 2PN accuracy or higher (for instance, the late inspiral phase of compact binaries detected by gravitational-wave observatories being a case in point ~\cite{Blanchet}) fall outside this approximation scheme and require the full ${}^{(4)}\!g_{00}$ to be worked out.

\section{Field Equations and Solutions}\label{sec:linearized_field_equations}

In Sec.~\ref{sec:brumberg} we showed that, in all physically viable ETG consistent with the screened regime $\Phi\sim\Psi$ (i.e.\ $\gamma\sim 1$), the linearized metric potentials $\Phi$, $\Psi$, and $Z_i$ already encode the complete 1PN dynamics of an $N$-body system. The purpose of the present section is therefore twofold: to cast the linearized STFOG field equations in the Standard Post-Newtonian (SPN) gauge, specialized to the point-particle energy-momentum tensor introduced below, and to solve the resulting fourth-order elliptic system by an explicit Green-function construction. This provides all post-Newtonian potentials entering the linearized line element \eqref{ds2-PN} for the full STFOG class, as well as for the NCSG subclass discussed in Sec.~\ref{sec:NCSG}.

\subsection{Effective masses and the linearized STFOG system}\label{sec:linearized-system}
The Lagrangian density $f(R,R_{\mu\nu}R^{\mu\nu},\phi)$ is Taylor-expanded around the Minkowski background $(R = 0,\,R_{\alpha\beta}R^{\alpha\beta} = 0,\,\phi = \phi^{(0)})$. The coefficients of this expansion, namely the partial derivatives evaluated at the background values, define the effective masses that govern the propagation of the additional degrees of freedom. In our notation, they are given by\,\footnote{In the Newtonian and post-Newtonian limits, one can consider the Lagrangian $f(R) = a\,R + b\,R^2 + c\,R_{\alpha\beta}R^{\alpha\beta}$. Then the masses \eqref{mass-full-notation} become $m_R^2 = -a/[2(3b+c)]$, $m_Y^2 = a/c$. For a correct interpretation of these quantities as real masses, one has to impose $a > 0$, $b < 0$ and $0 < c < -3b$.}
\begin{equation}\label{mass-full-notation}
\begin{aligned}
m_R^2 &= -\frac{f_R(0,0,\phi^{(0)})}{3\,f_{RR}(0,0,\phi^{(0)}) + 2\,f_Y(0,0,\phi^{(0)})}\,,\\[6pt]
m_Y^2 &= \frac{f_R(0,0,\phi^{(0)})}{f_Y(0,0,\phi^{(0)})}\,,\\[6pt]
m_\phi^2 &= -\frac{f_{\phi\phi}(0,0,\phi^{(0)})}{2\,\omega(\phi^{(0)})}\,.
\end{aligned}
\end{equation}
These masses are the inverses of the Compton wavelengths associated, respectively, with the scalar curvature mode (scalaron, mass $m_R$), the Ricci-tensor-squared mode (mass $m_Y$), and the scalar field $\phi$ (mass $m_\phi$). Setting\,\footnote{This amounts to redefining the gravitational coupling as $\mathcal{X} \to \mathcal{X}\,f_R(0,0,\phi^{(0)})$ and the non-minimal coupling as $f_{R\phi}(0,0,\phi^{(0)}) \to f_{R\phi}(0,0,\phi^{(0)})\,f_R(0,0,\phi^{(0)})$, so that $\mathcal{X}$ retains its standard value $8\pi G/c^{4}$.}
\begin{equation}\label{normalization}
f_R(0,0,\phi^{(0)}) = 1\,,\qquad \omega(\phi^{(0)}) = \tfrac{1}{2}\,,
\end{equation}
and, to lighten the notation hereafter, \textit{dropping the explicit background-value arguments} of all partial derivatives, i.e.\
\begin{equation}\label{notation-convention}
f_R \equiv f_R(0,0,\phi^{(0)})\,,\quad f_{RR} \equiv f_{RR}(0,0,\phi^{(0)})\,,\quad f_Y \equiv f_Y(0,0,\phi^{(0)})\,,\quad f_{R\phi} \equiv f_{R\phi}(0,0,\phi^{(0)})\,,\quad \text{etc.}\,,
\end{equation}
the effective masses simplify to
\begin{equation}\label{mass_definition}
m_R^2 = -\frac{1}{3f_{RR} + 2f_Y}\,,\qquad m_Y^2 = \frac{1}{f_Y}\,,\qquad m_\phi^2 = -f_{\phi\phi}\,.
\end{equation}

\subsubsection{Linearized field equations}
Applying the weak-field limit to \eqref{fieldequationSTFOG}, substituting the Ricci tensor components \eqref{ricci-potentials} and the energy-momentum tensor \eqref{Tmunu-covariant}, and using the effective masses \eqref{mass_definition} together with the normalization \eqref{normalization}, one obtains the complete system of linearized field equations for the potentials $\Phi$, $\Psi$, $Z_i$ and the scalar fields ${}^{(2)}\!R$, $\varphi$\,:
\begin{eqnarray}\label{PMfieldequationFOG3}\label{linearized-system}
&&(\nabla^2 - m_Y^2)\,\nabla^2\Phi + c^2\biggl[\frac{m_Y^2}{2} - \frac{m_R^2 + 2m_Y^2}{6\,m_R^2}\,\nabla^2\biggr]{}^{(2)}\!R + c^2m_Y^2\,f_{R\phi}\,\nabla^2\varphi
= -m_Y^2\,\mathcal{X}\sum_{a=1}^{N} m_a\,c^4\,\delta(\mathbf{x} - \mathbf{x}_a)\,,
\nonumber\\[6pt]
&&\biggl\{(\nabla^2 - m_Y^2)\,\nabla^2\Psi - c^2\biggl[\frac{m_Y^2}{2} - \frac{m_R^2 + 2m_Y^2}{6\,m_R^2}\,\nabla^2\biggr]{}^{(2)}\!R - c^2m_Y^2\,f_{R\phi}\,\nabla^2\varphi\biggr\}\,\delta_{ij}
\nonumber\\
&&\qquad + \biggl\{(\nabla^2 - m_Y^2)(\Psi - \Phi) + c^2\frac{m_R^2 - m_Y^2}{3\,m_R^2}\,{}^{(2)}\!R + c^2m_Y^2\,f_{R\phi}\,\varphi\biggr\}_{,ij} = 0\,,
\nonumber\\[6pt]
&&(\nabla^2 - m_Y^2)\,\nabla^2 Z_i + \biggl\{(\nabla^2 - m_Y^2)\,\Psi + 2c^2\biggl[\frac{m_R^2 - m_Y^2}{3\,m_R^2}\,{}^{(2)}\!R + m_Y^2\,f_{R\phi}\,\varphi\biggr]\biggr\}_{,0i}
\nonumber\\
&&\qquad\qquad\qquad\qquad\qquad\qquad\qquad\qquad\qquad
= 2\,m_Y^2\,\mathcal{X}\sum_{a=1}^{N} m_a\,c^4\,v_i^a\,\delta(\mathbf{x} - \mathbf{x}_a)\,,
\nonumber\\[6pt]
&&(\nabla^2 - m_R^2)\,{}^{(2)}\!R - 3\,m_R^2\,f_{R\phi}\,\nabla^2\varphi
= m_R^2\,\mathcal{X}\sum_{a=1}^{N} m_a\,c^2\,\delta(\mathbf{x} - \mathbf{x}_a)\,,
\nonumber\\[6pt]
&&(\nabla^2 - m_\phi^2)\,\varphi + f_{R\phi}\,{}^{(2)}\!R = 0\,.
\end{eqnarray}

System \eqref{PMfieldequationFOG3} consists of five coupled, fourth-order (in the first three rows) and second-order (in the last two) partial differential equations, where $\nabla^2$ and $(\nabla^2 - m_Y^2)$ are the spatial Laplacian and Klein--Gordon operators, respectively. The first three equations govern the gravitational potentials $\Phi$, $\Psi$, $Z_i$ and involve the scalar fields ${}^{(2)}\!R$ and $\varphi$ as sources; the last two equations form a self-contained coupled subsystem for ${}^{(2)}\!R$ and $\varphi$.
The system \eqref{PMfieldequationFOG3} contains products of the flat-space Laplacian $\nabla^2$ and of massive Klein--Gordon operators $(\nabla^2 - m^2)$, with $m \in \{m_R, m_Y, m_\phi\}$. Its inversion is carried out by the standard Green-function decomposition for composed elliptic operators, whose explicit construction and the full list of Yukawa-convolution integrals required at each step are detailed in Refs.~\cite{FOG_CGL,FOG1,FOG2,FOGST,TedescoOrbital2023}.

%------------------------------------------------------

%------------------------------------------------------
\subsection{Scalar field solutions for \texorpdfstring{$\varphi$ and ${}^{(2)}\!R$}{varphi and R(2)}}\label{sec:scalar-solutions}

Equations \eqref{PMfieldequationFOG3}$_4$ and \eqref{PMfieldequationFOG3}$_5$ form a coupled subsystem for the Ricci scalar perturbation ${}^{(2)}\!R$ and the scalar field $\varphi$, decoupled from the gravitational potentials. Diagonalisation yields the mass eigenvalues $m_\pm^2 = m_R^2\,\omega_\pm$, with
\begin{equation}\label{omega-pm}
\omega_\pm = \frac{1 - \xi + \eta^2 \pm \sqrt{(1 - \xi + \eta^2)^2 - 4\eta^2}}{2}\,,
\end{equation}
$\xi = 3\,f_{R\phi}^2$, $\eta = m_\phi/m_R$, and the reality condition $(\eta - 1)^2 - \xi > 0$ \cite{FOG_CGL,Stabile2014,FOG1,FOG2}. Both eigenmodes satisfy Klein--Gordon equations with masses $m_\pm$, and their solutions follow from the standard Green-function inversion (see Refs.~\cite{FOG_CGL,FOG1,FOG2,FOGST,TedescoOrbital2023}). For the $N$-body source one obtains
\begin{equation}\label{scalar-field-solutions}
\begin{aligned}
\varphi(\mathbf{x}) &=
\sqrt{\frac{\xi}{3}}\;\frac{G}{c^2}
\sum_{a=1}^{N}
\frac{m_a}{|\mathbf{x} - \mathbf{x}_a|}
\frac{e^{-m_+ |\mathbf{x} - \mathbf{x}_a|} - e^{-m_- |\mathbf{x} - \mathbf{x}_a|}}{\omega_+ - \omega_-}\,,\\[6pt]
{}^{(2)}\!R(\mathbf{x}) &=
-m_R^2\;\frac{G}{c^2}
\sum_{a=1}^{N}
\frac{m_a}{|\mathbf{x} - \mathbf{x}_a|}
\frac{(\omega_+ - \eta^2)\,e^{-m_+ |\mathbf{x} - \mathbf{x}_a|} - (\omega_- - \eta^2)\,e^{-m_- |\mathbf{x} - \mathbf{x}_a|}}{\omega_+ - \omega_-}\,.
\end{aligned}
\end{equation}
Both fields decay exponentially at infinity with Compton wavelengths $m_\pm^{-1}$; the elementary one-body solution agrees with the harmonic-gauge computation of Ref.~\cite{FOGST}, consistently with the gauge invariance of the scalar sector.

%------------------------------------------------------
\subsection{Gravitational potentials \texorpdfstring{$\Phi$ and $\Psi$}{Phi and Psi}}\label{sec:Phi-Psi-solutions}

We turn to the gravitational potentials. $\Phi$ follows from Eq.~\eqref{PMfieldequationFOG3}$_1$ after eliminating ${}^{(2)}\!R$ and $\varphi$ via \eqref{scalar-field-solutions} and applying the Green-function inversion mentioned above. After evaluating term by term the Yukawa convolutions of the profiles $e^{-m_\pm |\mathbf{x} - \mathbf{x}_a|}/|\mathbf{x} - \mathbf{x}_a|$ generated by ${}^{(2)}\!R$ and $\varphi$, and combining the result with the point-source contribution, the elementary solution generated by the body $a$ is reorganized in the compact form
\begin{equation}\label{Phi-point}
\Phi(\mathbf{x}) =
-G\sum_{a=1}^{N}\frac{m_a}{|\mathbf{x} - \mathbf{x}_a|}
\biggl\{1 + g(\xi,\eta)\,e^{-m_+|\mathbf{x} - \mathbf{x}_a|}
+ \Bigl(\tfrac{1}{3} - g(\xi,\eta)\Bigr)e^{-m_-|\mathbf{x} - \mathbf{x}_a|}
- \tfrac{4}{3}e^{-m_Y|\mathbf{x} - \mathbf{x}_a|}\biggr\}\,,
\end{equation}
with the structure function
\begin{equation}\label{g-function}
g(\xi,\eta) = \frac{1 - \eta^2 + \xi + \sqrt{\eta^4 + (\xi - 1)^2 - 2\eta^2(\xi + 1)}}{6\,\sqrt{\eta^4 + (\xi - 1)^2 - 2\eta^2(\xi + 1)}}\,.
\end{equation}
The potential $\Psi$ is determined from \eqref{PMfieldequationFOG3}$_2$ by imposing separately the traceless ($\{\ldots\}_{,ij}=0$) and the trace ($\{\ldots\}\delta_{ij}=0$) parts; one obtains 
\begin{equation}\label{Psi-point}
\Psi(\mathbf{x}) =
-G\sum_{a=1}^{N}\frac{m_a}{|\mathbf{x} - \mathbf{x}_a|}
\biggl\{1 - g(\xi,\eta)\,e^{-m_+|\mathbf{x} - \mathbf{x}_a|}
- \Bigl(\tfrac{1}{3} - g(\xi,\eta)\Bigr)e^{-m_-|\mathbf{x} - \mathbf{x}_a|}
- \tfrac{2}{3}e^{-m_Y|\mathbf{x} - \mathbf{x}_a|}\biggr\}\,.
\end{equation}
We see that by the linearity of \eqref{PMfieldequationFOG3}, $\Phi$ splits as a linear combination of the $f(R,\phi)$ and 
$f(R, R_{\alpha\beta}R^{\alpha\beta})$ contributions, a property 
inherited from the absence of a $Y\,\phi$ cross-coupling 
($f_{Y\phi}|_0 = 0$) in the expansion of \eqref{FOGaction}. For $f_Y\to 0$ ($m_Y\to\infty$), the $e^{-m_Y|\mathbf{x}|}$ mode is suppressed and the $f(R,\phi)$ potentials of Ref.~\cite{FOGST} are recovered; in the GR limit ($m_R,m_Y,m_\phi\to\infty$) all Yukawa corrections vanish and $\Phi=\Psi=-GM/|\mathbf{x}|$.
For later use in the $N$-body Lagrangian of Sec.~\ref{sec:relativistic_dynamics}, we introduce the compact notation
\begin{equation}\label{zeta-def}
\Phi(\mathbf{x}) = -G\sum_{a=1}^{N}\frac{m_a}{|\mathbf{x} - \mathbf{x}_a|}\bigl\{1 + \zeta_a(\mathbf{x})\bigr\}\,,\qquad
\Psi(\mathbf{x}) = -G\sum_{a=1}^{N}\frac{m_a}{|\mathbf{x} - \mathbf{x}_a|}\bigl\{1 - \tilde{\zeta}_a(\mathbf{x})\bigr\}\,,
\end{equation}
with
\begin{equation}\label{zeta-explicit}
\begin{aligned}
\zeta_a(\mathbf{x}) &\equiv g(\xi,\eta)\,e^{-m_+ |\mathbf{x} - \mathbf{x}_a|} + \bigl[\tfrac{1}{3} - g(\xi,\eta)\bigr]\,e^{-m_- |\mathbf{x} - \mathbf{x}_a|} - \tfrac{4}{3}\,e^{-m_Y |\mathbf{x} - \mathbf{x}_a|}\,,\\
\tilde{\zeta}_a(\mathbf{x}) &\equiv g(\xi,\eta)\,e^{-m_+ |\mathbf{x} - \mathbf{x}_a|} + \bigl[\tfrac{1}{3} - g(\xi,\eta)\bigr]\,e^{-m_- |\mathbf{x} - \mathbf{x}_a|} + \tfrac{2}{3}\,e^{-m_Y |\mathbf{x} - \mathbf{x}_a|}\,.
\end{aligned}
\end{equation}
We note that solutions \eqref{Phi-point} and \eqref{Psi-point} consistently recover the known point-like source potentials in the full STFOG notation
of Refs.~\cite{FOG1,FOG2,FOG_CGL,FOGST}; the harmonic-gauge post-Newtonian treatment of the $f(R)$ subclass is discussed in Ref.~\cite{PRD1}. 

The novelty of the present construction is its use for post-Newtonian $N$-body systems in the Standard Post-Newtonian gauge 
\eqref{SPN-gauge}, which is the natural choice for relativistic
celestial mechanics~\cite{Brumberg,BrumbergBook1991,Kopeikin2011,Soffel} and differs from the
harmonic representation in the gravitomagnetic sector. In fact, as we will see in Sec.~\ref{sec:Zi-solution}, while $\Phi$
and $\Psi$ are unchanged at the Newtonian-like order retained here,
the metric cross-term is the SPN combination
$Z_i = A_i - X_{,0i}$.

%------------------------------------------------------
\subsection{Gravitomagnetic potentials \texorpdfstring{$Z_i$}{Zi}, \texorpdfstring{$A_i$}{Ai}, and \texorpdfstring{$X$}{X}}\label{sec:Zi-solution}

The $0i$-equation \eqref{PMfieldequationFOG3}$_3$ determines the gravitomagnetic potential $Z_i = A_i - X_{,0i}$, where $A_i$ is the vector potential and $X$ the superpotential satisfying $\nabla^2 X = \Phi$. The traceless condition $\{\ldots\}_{,ij} = 0$ in Eq.~\eqref{PMfieldequationFOG3}$_2$, combined with the SPN gauge relation \eqref{SPN-gauge-explicit}, yields the additional constraint $(\nabla^2 - m_Y^2)\,\Phi_{,0i} = 0$ at the relevant order. Strictly speaking, this reduction is understood in the regime 
$(\Phi \sim \Psi)$, consistently with the approximation adopted in Sec.~3; 
terms proportional to \((\Phi-\Psi)_{,0i}\) are therefore beyond the retained 
accuracy. Consequently, for a system of $N$ bodies with masses $m_a$ and velocities $v^a_i$, Eq.~\eqref{PMfieldequationFOG3}$_3$ simplifies to
\begin{equation}\label{Zi-equation}
(\nabla^2 - m_Y^2)\,\nabla^2\,(Z_i + X_{,0i})
=
2\,m_Y^2\,\mathcal{X}
\sum_{a=1}^{N}m_a\,c^4\,v_i^a\,
\delta(\mathbf{x}-\mathbf{x}_a)\,.
\end{equation}
Applying the same Green-function inversion to the right-hand side of this equation, we obtain
\begin{equation}\label{Zi-solution}
\begin{aligned}
Z_i(\mathbf{x}) = 4G\sum_{a=1}^{N}\frac{m_a}{|\mathbf{x} - \mathbf{x}_a|}\,v_i^a - 4G\sum_{a=1}^{N}\frac{m_a\,e^{-m_Y|\mathbf{x} - \mathbf{x}_a|}}{|\mathbf{x} - \mathbf{x}_a|}\,v_i^a - X_{,0i}(\mathbf{x})\,,
\end{aligned}
\end{equation}
where the first two terms define the vector potential
\begin{equation}\label{Ai-point}
A_i(\mathbf{x}) =
4G\sum_{a=1}^{N}\frac{m_a}{|\mathbf{x} - \mathbf{x}_a|}\bigl(1-e^{-m_Y|\mathbf{x} - \mathbf{x}_a|}\bigr)\,v_i^a\,.
\end{equation}
The first contribution is the GR result and the second is the massive-mode correction induced by the $R_{\alpha\beta}R^{\alpha\beta}$ invariant. To complete the solution, it remains to solve the equation
\begin{equation}\label{super-pot_equation}
\nabla^2 X = \Phi\,,
\end{equation}
with $\Phi$ given by Eq.~\eqref{zeta-def}. The full $N$-body superpotential reads
\begin{equation}\label{Superpotential-result}
\begin{aligned}
X(\mathbf{x})
=
-G\sum_{a=1}^{N}m_a
\biggl\{
\frac{|\mathbf{x} - \mathbf{x}_a|}{2}
-g(\xi,\eta)\,
\frac{1-e^{-m_+|\mathbf{x} - \mathbf{x}_a|}}{m_+^2\,|\mathbf{x} - \mathbf{x}_a|} 
-\bigl(\tfrac{1}{3}-g(\xi,\eta)\bigr)\,
\frac{1-e^{-m_-|\mathbf{x} - \mathbf{x}_a|}}{m_-^2\,|\mathbf{x} - \mathbf{x}_a|}
+\frac{4}{3}\,
\frac{1-e^{-m_Y|\mathbf{x} - \mathbf{x}_a|}}{m_Y^2\,|\mathbf{x} - \mathbf{x}_a|}
\biggr\}\,. 
\end{aligned}
\end{equation}
The full gravitomagnetic potential is thus
\begin{equation}\label{Zi-Nbody}
Z_i(\mathbf{x})=
A_i(\mathbf{x})-X_{,0i}(\mathbf{x})\,,
\end{equation}
which extends the GR result by incorporating the Yukawa-type modifications to both the vector potential and the superpotential. We remark that the time and spatial derivatives of $X$ in $X_{,0i}$ generate additional post-Newtonian contributions absent in harmonic gauge (where $Z_i \equiv A_i$), but the physical content of the equations of motion is gauge-independent \cite{Brumberg,BrumbergBook1991,Brumberg2,Soffel}.

%------------------------------------------------------
\subsection{Compact summary and theory limits}\label{sec:summary-solutions}

The complete linearized metric for the STFOG class, generated by the $N$-body point-particle source \eqref{Tmunu-covariant} in the SPN gauge, is given by Eqs.~\eqref{potentials-def} and \eqref{ds2-PN} with the potentials \eqref{zeta-def}, \eqref{Ai-point}, \eqref{Superpotential-result}, and \eqref{Zi-Nbody}. The corresponding elementary solutions entering the variational $N$-body construction of Sec.~\ref{sec:relativistic_dynamics} are given by Eqs.~\eqref{Phi-point}, \eqref{Psi-point}, \eqref{Ai-point}, and \eqref{Superpotential-result}. The scalar-sector parameters $\xi$, $\eta$, $\omega_\pm$, and $g(\xi,\eta)$ are defined in Eqs.~\eqref{omega-pm}--\eqref{g-function}, and the effective masses are given by Eqs.~\eqref{mass-full-notation}--\eqref{mass_definition}. Table~\ref{tab:ETG-cases} summarizes the effective masses and their limiting behavior for the principal subclasses of ETG.

\renewcommand{\arraystretch}{1.8}
\begin{table}[ht]
\centering
\begin{tabular}{@{\hspace{0.1cm}}l@{\hspace{0.3cm}}l@{\hspace{0.3cm}}c@{\hspace{0.3cm}}c@{\hspace{0.3cm}}c@{\hspace{0.3cm}}c@{\hspace{0.3cm}}c@{\hspace{0.1cm}}}
\specialrule{1.5pt}{0pt}{0pt}
\textbf{Case} & \textbf{Theory} & $m_R^2$ & $m_Y^2$ & $m_\phi^2$ & $m_+^2$ & $m_-^2$ \\
\midrule
A & $f(R)$ & $-\frac{f_R}{3f_{RR}}$ & $\infty$ & $0$ & $m_R^2$ & $\infty$ \\[4pt]
B & $f(R,R_{\alpha\beta}R^{\alpha\beta})$ & $-\frac{f_R}{3f_{RR}+2f_Y}$ & $\frac{f_R}{f_Y}$ & $0$ & $m_R^2$ & $\infty$ \\[4pt]
C & $f(R,\phi)+\omega\,\phi_{;\alpha}\phi^{;\alpha}$ & $-\frac{f_R}{3f_{RR}}$ & $\infty$ & $-\frac{f_{\phi\phi}}{2\omega}$ & $m_R^2\omega_+$ & $m_R^2\omega_-$ \\[4pt]
D & $f(R,R_{\alpha\beta}R^{\alpha\beta},\phi)+\omega(\phi)\phi_{;\alpha}\phi^{;\alpha}$  & $-\frac{f_R}{3f_{RR}+2f_Y}$ & $\frac{f_R}{f_Y}$ & $-\frac{f_{\phi\phi}}{2\omega}$ & $m_R^2\omega_+$ & $m_R^2\omega_-$ \\
\specialrule{1.5pt}{0pt}{0pt}
\end{tabular}
\caption{Effective masses for the principal subclasses of ETG, with $f_R = 1$ and $\omega(\phi^{(0)}) = 1/2$. Case~D is the full STFOG; cases A--C are recovered in the appropriate limits (see \cite{FOG_CGL} for details). The condition $(\eta - 1)^2 - \xi > 0$ ensures real masses in cases~C and~D. For cases A and B the scalar field is absent: the entries $m_\phi^2$ and $m_-^2$ denote the decoupled scalar sector, whose Yukawa contribution vanishes because the coefficient $(1/3-g)$ is zero at $\xi=\eta=0$.}
\label{tab:ETG-cases}
\end{table}

\subsubsection{NCSG solutions}\label{sec:NCSG-solutions}

As discussed in Sec.~\ref{sec:NCSG}, the NCSG sector follows by solving the field equation \eqref{NCSG-FE} directly in the SPN gauge. Since the bracket in \eqref{NCSG-FE} is the (traceless) Bach tensor, the trace of \eqref{NCSG-FE} gives $R = -\mathcal{X}\,T$, i.e. ${}^{(2)}\!R = 0$ in vacuum; using the SPN expression for the Ricci scalar \eqref{Ricci-scalar-SPN-2}, this relates the spatial potential to the temporal one through $\nabla^2\Psi = 1/2 \,\nabla^2\Phi$ outside the sources. The $00$-equation fixes $\Phi$, which equals the regular $m_R\to\infty$ limit of \eqref{Phi-point} with $m_Y=\beta$ (the scalar modes decouple, $\xi=\eta=0$, $g(0,0)=1/3$, and the $m_\pm$ Yukawa terms drop out), so that the temporal Yukawa coefficient is $4/3$; the trace relation then fixes the spatial Yukawa coefficient to one half of it, $2/3$. Equivalently, both potentials are the regular $m_Y$-block limit of \eqref{Phi-point}--\eqref{Psi-point}; they agree with the harmonic-coordinate calculation of Ref.~\cite{Lambiase:2013dai} for $\Phi$ and for the gravitomagnetic sector. In the weak-field limit, for a system of $N$ point particles the potentials read
\begin{equation}\label{NCSG-potentials}
\begin{aligned}
\Phi_{\text{\tiny NCSG}}(\mathbf{x})
&=
-G\sum_{a=1}^{N}\frac{m_a}{|\mathbf{x} - \mathbf{x}_a|}
\biggl\{1 - \tfrac{4}{3}\,e^{-\beta |\mathbf{x} - \mathbf{x}_a|}\biggr\}\,,\\[4pt]
\Psi_{\text{\tiny NCSG}}(\mathbf{x})
&=
-G\sum_{a=1}^{N}\frac{m_a}{|\mathbf{x} - \mathbf{x}_a|}
\biggl\{1 - \tfrac{2}{3}\,e^{-\beta |\mathbf{x} - \mathbf{x}_a|}\biggr\}\,.
\end{aligned}
\end{equation}
The vector potential and the superpotential are
\begin{equation}\label{NCSG-Ai-X}
\begin{aligned}
A_i^{\text{\tiny NCSG}}(\mathbf{x})
&=
4G\sum_{a=1}^{N}
\frac{m_a}{|\mathbf{x} - \mathbf{x}_a|}
\bigl\{1 - e^{-\beta |\mathbf{x} - \mathbf{x}_a|}\bigr\}\,v_i^a\,,\\[4pt]
X^{\text{\tiny NCSG}}(\mathbf{x})
&=
-\frac{G}{2}\sum_{a=1}^{N}m_a |\mathbf{x} - \mathbf{x}_a|
-\frac{4G}{3\beta^2}
\sum_{a=1}^{N}\frac{m_a}{|\mathbf{x} - \mathbf{x}_a|}
\biggl\{1 - e^{-\beta |\mathbf{x} - \mathbf{x}_a|} \biggr\}\,,
\end{aligned}
\end{equation}
and the complete SPN gravitomagnetic potential is
\begin{equation}\label{NCSG-Zi}
Z_i^{\text{\tiny NCSG}}(\mathbf{x})
=
A_i^{\text{\tiny NCSG}}(\mathbf{x})
-
X^{\text{\tiny NCSG}}_{,0i}(\mathbf{x})\,.
\end{equation}
Equations~\eqref{NCSG-potentials}--\eqref{NCSG-Zi} reproduce the NCSG solution in the weak-field limit for $g_{00}$ and $g_{ij}$ at the level of the elementary point-like sources and give directly their $N$-body superposition. They also make explicit the SPN-gauge completion of the gravitomagnetic sector, which differs from the harmonic-gauge form precisely by the additional $-X_{,0i}$ contribution \cite{Lambiase:2013dai,NaefJetzer2010,TedescoOrbital2023}. The spatial Yukawa coefficient $2/3$ in $\Psi_{\text{\tiny NCSG}}$ is exactly one half of the temporal coefficient $4/3$ in $\Phi_{\text{\tiny NCSG}}$, as enforced by the tracelessness of the Bach tensor ($R=0$ in vacuum); equivalently, it is the regular $m_Y$-block limit of \eqref{Psi-point}.\footnote{The harmonic-gauge analysis of Ref.~\cite{Lambiase:2013dai} quotes a spatial coefficient $5/9$ for $g_{ij}$. That value is the orbit-averaged ($\langle x^i x^j\rangle = \tfrac{r^2}{3}\,\delta^{ij}$) projection of the anisotropic solution, appropriate for the gyroscope-precession observable considered there; it is not the isotropic line-element potential entering \eqref{ds2-PN}, whose coefficient is fixed to $2/3$ by $R=0$.} These potentials, inserted into the line element \eqref{ds2-PN}, provide the complete 1PN metric for NCSG needed in the next section.

\section{Post-Newtonian Equations of Motion for the Relativistic \texorpdfstring{$N$}{N}-body System in STFOG}\label{sec:relativistic_dynamics}

\subsection{Variational principle of the 1PN dynamics for STFOG}\label{sec:variational}

In this section we offer an independent proof that ${}^{(4)}\!g_{00}$ is nonessential at 1PN under the hypothesis $\Phi \sim \Psi$, and derive the relativistic Lagrangian from which the post-Newtonian equations of motion for $N$ gravitating bodies follow. Both results arise naturally from the variational principle in the Infeld--Brumberg sense: the post-Newtonian metric ansatz of Sec.~\ref{sec:PNformalism} is substituted into the action with the metric components already written in terms of the post-Newtonian potentials, and the reduced particle Lagrangian is then isolated once the field equations are used inside the spatial integrals. Each body is described by a timelike worldline, and the role of ${}^{(4)}\!g_{00}$ can be read off directly at the level of the action, without having to solve first for the full nonlinear metric.

For STFOG it is convenient to separate from the outset the Einstein--Hilbert contribution from the genuinely extra-curvature sector. The Fock--Brumberg decomposition of $\sqrt{-g}\,R$ applies only to the Einstein--Hilbert part; the higher-curvature terms, on the other hand, have to be organized through an explicit post-Newtonian counting. We therefore decompose the STFOG action \eqref{FOGaction} as 
\begin{equation}\label{S-decomposition}
\mathcal{S} = \mathcal{S}_g + \mathcal{S}_g^{\,\text{extra}} + \mathcal{S}_m\,.
\end{equation}
Hereafter, $d^4x = c\,dt\,d^3\mathbf{x}$. The splitting \eqref{S-decomposition} is understood after expanding the STFOG Lagrangian density around the Minkowski background $(R=0,\;R_{\mu\nu}R^{\mu\nu}=0,\;\phi=\phi^{(0)})$, with
\begin{equation}\label{STFOG-background-conditions}
f(0,0,\phi^{(0)})=0\,,\qquad
f_R(0,0,\phi^{(0)})=1\,,\qquad
f_\phi(0,0,\phi^{(0)})=0\,,\qquad
\omega(\phi^{(0)})=\frac12\,.
\end{equation}
In the post-Newtonian regime one has, and we continue to denote by $Y$ the Ricci-tensor invariant $R_{\mu\nu}R^{\mu\nu}$,
\begin{equation}\label{STFOG-PN-bookkeeping}
R = O(\varepsilon^{2})\,,\qquad
R_{\mu\nu}=O(\varepsilon^{2})\,,\qquad
\varphi = O(\varepsilon^{2})\,,\qquad
Y = R_{\mu\nu}R^{\mu\nu}=O(\varepsilon^{4})\,,
\end{equation}
so that, up to the accuracy relevant for the 1PN dynamics, the function $f$ contributes only through
\begin{equation}\label{STFOG-Taylor-1PN}
f(R,Y,\phi)=
R+\frac{f_{RR}}{2}\,R^2+f_Y\,R_{\mu\nu}R^{\mu\nu}+f_{R\phi}\,R\,\varphi+\frac{f_{\phi\phi}}{2}\,\varphi^2+O(\varepsilon^{6})\,.
\end{equation}
In this form, $\mathcal{S}_g$ contains only the Einstein--Hilbert piece, while $\mathcal{S}_g^{\,\textnormal{extra}}$ collects precisely the lowest extra-curvature and scalar terms relevant at 1PN order.

\medskip
\noindent\textit{1) Gravitational action.} The Ricci scalar admits the identity \cite{Brumberg2,Fock}
\begin{equation}\label{Ricci-decomposition}
\sqrt{-g}\;R = \Bigl[\sqrt{-g}\;\bigl(g^{\mu\nu}\,\Gamma^\alpha_{\alpha\nu} - g^{\alpha\nu}\,\Gamma^\mu_{\alpha\nu}\bigr)\Bigr]_{,\mu} - \sqrt{-g}\;\bar{R}\,,
\end{equation}
where $\bar{R} \equiv g^{\mu\nu}\bigl(\Gamma^\alpha_{\beta\nu}\Gamma^\beta_{\alpha\mu} - \Gamma^\alpha_{\mu\nu}\Gamma^\beta_{\alpha\beta}\bigr)$ depends on $g_{\mu\nu}$ and its first derivatives only. Equation~\eqref{Ricci-decomposition} is used here exclusively for the Einstein--Hilbert sector. Since the first term is a total divergence, it does not contribute when $\delta g_{\mu\nu}\big|_{\partial V}=0$, and, after discarding the boundary contribution, the Einstein--Hilbert action can be written as
\begin{equation}\label{Sg-def}
\mathcal{S}_g = -\frac{c^4}{16\pi G}\int\!\sqrt{-g}\;\bar{R}\,dt\,d^3\mathbf{x}\,,\qquad\Longrightarrow\qquad L_g = - \frac{c^4}{16\pi G} \!\int\!\sqrt{-g}\;\bar{R}\;d^3\mathbf{x}\,.
\end{equation}
A crucial point for the higher-curvature extension is that the same manipulation cannot be transferred to $R^2$ before the post-Newtonian truncation. Indeed, if we define
\begin{equation}\label{Rpartial-def}
R_*= \frac{1}{\sqrt{-g}}\,
\partial_\mu\!\Bigl[\sqrt{-g}\,\bigl(g^{\mu\nu}\Gamma^\alpha_{\alpha\nu}-g^{\alpha\nu}\Gamma^\mu_{\alpha\nu}\bigr)\Bigr]\,,
\qquad\Rightarrow\qquad
R=R_*-\bar{R}\,,
\end{equation}
then
\begin{equation}\label{R2-split-warning}
R^2 = R_*^{\,2} - 2\,R_*\,\bar{R} + \bar{R}^{\,2}\,.
\end{equation}
Now $R_*=O(\varepsilon^{2})$, whereas $\bar{R}=O(\varepsilon^{4})$, so that
\begin{equation}\label{R2-orders-warning}
R_*^{\,2}=O(\varepsilon^{4})\,,\qquad
R_{*}\bar{R}=O(\varepsilon^{6})\,,\qquad
\bar{R}^{\,2}=O(\varepsilon^{8})\,.
\end{equation}
Therefore the boundary-term elimination leading to \eqref{Sg-def} is specific to the Einstein--Hilbert sector: once squared, the divergence part contributes already at 1PN order and must be retained in $\mathcal{S}_g^{\,\textnormal{extra}}$.
Carrying out the post-Newtonian expansion of the field Lagrangian $L_g$ for a system of $N$ point masses (using the Poisson equation in the spatial integrals), the linear-sector contribution can be written as
\begin{equation}\label{Lg-expanded}
L_g = \frac{1}{2}\sum_a m_a\biggl[\Phi^{(a)} - \frac{\bigl(\Phi^{(a)}\bigr)^2}{c^2} + \frac{Z_i^{(a)}\,v_a^i}{c^2} + c^2\,{}^{(4)}\!g_{00}^{\,(a)}\biggr] + O(\varepsilon^{4})\,.
\end{equation}
We stress the presence, within $L_g$, of the explicit field contribution
$1/2 \sum_a m_a\,c^2\;{}^{(4)}\!g_{00}^{\,(a)}$.
While the extra-curvature and scalar field corrections to the gravitational action beyond GR give
\begin{equation}\label{Sextra-def}
\begin{aligned}
\mathcal{S}_g^{\,\text{extra}}
&= \frac{c^4}{16\pi G}\int\!\sqrt{-g}\;\biggl[\frac{f_{RR}}{2}\,R^2 + f_Y\,R_{\alpha\beta}R^{\alpha\beta} + f_{R\phi}\,R\,\varphi
+ \frac{f_{\phi\phi}}{2}\,\varphi^2 + \frac{1}{2}\,\varphi_{;\alpha}\varphi^{;\alpha}\biggr]\,dt\,d^3\mathbf{x}\,,\\
\Longrightarrow\qquad
L_g^{\,\text{extra}}
&= \frac{c^4}{16\pi G}\!\int\!\sqrt{-g}\;\biggl[\frac{f_{RR}}{2}\,R^2 + f_Y\,R_{\alpha\beta}R^{\alpha\beta} + f_{R\phi}\,R\,\varphi
+ \frac{f_{\phi\phi}}{2}\,\varphi^2 + \frac{1}{2}\,\varphi_{;\alpha}\varphi^{;\alpha}\biggr]d^3\mathbf{x}\,.
\end{aligned}
\end{equation}

\medskip
\noindent\textit{2) Matter action.} For a system of $N$ point particles, the matter action reads
\begin{equation}\label{Sm-def}
\mathcal{S}_m = -\sum_{a=1}^{N}m_a\,c\!\int\!ds_a\,,\qquad\Longrightarrow\qquad L_m = -\sum_{a=1}^{N}m_a\,c^2\;\frac{ds_a}{c\,dt}\,.
\end{equation}
Utilizing the 1PN metric components $g_{00} = 1 + 2\Phi/c^2 + {}^{(4)}\!g_{00}$, $g_{0i} = Z_i/c^3$, $g_{ij} = -(1 - 2\Psi/c^2)\,\delta_{ij}$,
expanding $ds/(c\,dt)$ up to $\mathcal{O}(\varepsilon^{4})$, and assuming $\Phi\sim\Psi$ (Sec.~\ref{sec:brumberg}), the matter Lagrangian reads (up to an irrelevant rest-energy constant)
\begin{equation}\label{Lm-expanded}
L_m = \sum_a\biggl[\frac{1}{2}\,m_a\,v_a^2 - m_a\,\Phi^{(a)}\biggr] + \frac{1}{c^2}\sum_a\biggl[\frac{1}{8}\,m_a\,v_a^4 - m_a\,Z_i^{(a)}\,v^i_a - \frac{3}{2}\,m_a\,\Phi^{(a)}\,v_a^2 + \frac{1}{2}\,m_a\,\bigl(\Phi^{(a)}\bigr)^2\biggr] - \frac{1}{2}\sum_a m_a\,c^2\;{}^{(4)}\!g_{00}^{\,(a)}\,.
\end{equation}
where $\Phi^{(a)} = \sum_{b\neq a}\Phi_b(\mathbf{x}_a)$, $Z_i^{(a)} = \sum_{b\neq a}(Z_i)_b(\mathbf{x}_a)$, and we recall that $Z_i = A_i - \partial^2 X/(\partial t\,\partial x^i)$. The first line is the Newtonian kinetic and potential energy; the second line contains the 1PN corrections, the $v^4$ term, the gravitomagnetic coupling via $Z_i$, the kinetic potential cross term, and the $\Phi^2$ nonlinearity. The last line exhibits the explicit ${}^{(4)}\!g_{00}$ contribution from the expansion of $g_{00}$.

\medskip
The total Lagrangian is therefore $L = L_g + L_g^{\,\text{extra}} + L_m$, and the variational principle $\delta\!\int\!L\,dt = 0$ yields the field equations \eqref{fieldequationSTFOG}, \eqref{traceSTFOG} upon varying with respect to $g_{\mu\nu}$ and $\phi$. Combining \eqref{Lg-expanded} and \eqref{Lm-expanded}, the ${}^{(4)}\!g_{00}$ terms cancel exactly, and the remaining field-sector contributions modify the velocity-dependent and gravitomagnetic terms in $L_m$ \cite{Brumberg2}. The combined single-body Lagrangian $L_a = (L_g + L_m)_a$, after this cancellation and the rearrangement of the field-sector contributions, takes the form\footnote{The factor $1/2$ multiplying the gravitomagnetic bracket $[A_i - \partial^2 X/(\partial t\,\partial x^i)]$ arises from the combination of the matter and field sectors: $L_m$ alone contributes coefficient $1$, while $L_g$ contributes $-1/2$, yielding $1/2$ in the total \cite{Brumberg2}.}
\begin{equation}\label{variational-Lagrangian}
L = \frac{1}{2}\,m_a\,v_a^2 + \frac{m_a\,v_a^4}{8\,c^2} - m_a\,c^2\,\biggl(\frac{\Phi^{(a)}}{c^2} + \biggl[A_i^{(a)} - \frac{\partial^2 X^{(a)}}{\partial t\,\partial x^i}\biggr]\frac{v_a^i}{2\,c^4} + \frac{3\,\Phi^{(a)}}{2}\;\frac{v_a^2}{c^4} + \frac{\bigl(\Phi^{(a)}\bigr)^2}{2\,c^4}\biggr)\,,
\end{equation}
where $\Phi^{(a)} = \sum_{b\neq a}\Phi_b(\mathbf{x}_a)$, $A_i^{(a)} = \sum_{b\neq a}(A_i)_b(\mathbf{x}_a)$, $X^{(a)} = \sum_{b\neq a}X_b(\mathbf{x}_a)$, with the potentials given by Eqs.~\eqref{Phi-point}, \eqref{Ai-point}, and \eqref{Superpotential-result}. Equation~\eqref{variational-Lagrangian} shows that the 1PN Lagrangian obtained from the Einstein--Hilbert plus matter sectors does not depend on the nonlinear metric component ${}^{(4)}\!g_{00}$. 

At this stage, however, one should distinguish two logically different mechanisms. In the Einstein--Hilbert plus matter sectors, the disappearance of ${}^{(4)}\!g_{00}$ follows from an exact cancellation between the gravitational and material contributions. In the genuinely higher-curvature sector, instead, its absence at 1PN order will follow from post-Newtonian counting rather than from a second cancellation. This point is established in the next subsection.

%----------------------------------------------------------------------
\subsection{Sufficiency of the linearized metric}
The variational argument must now be completed for the full STFOG class. Two distinct mechanisms are in fact at work. In the Einstein--Hilbert sector the Lagrangian loses its dependence on ${}^{(4)}\!g_{00}$ via the Brumberg cancellation recalled above; in the genuinely higher-curvature sector, on the other hand, there is no companion term against which ${}^{(4)}\!g_{00}$ could cancel: it simply does not enter at 1PN order because the relevant invariants are already quadratic in the linearized curvatures and in the scalar perturbation.

To make this explicit, we expand the curvature invariants as
\begin{equation}\label{curvature-1PN-expansion}
R={}^{(2)}\!R+{}^{(4)}\!R+O(\varepsilon^{6})\,,\qquad
R_{\mu\nu}={}^{(2)}\!R_{\mu\nu}+{}^{(4)}\!R_{\mu\nu}+O(\varepsilon^{6})\,,
\end{equation}
where ${}^{(4)}\!R$ and ${}^{(4)}\!R_{\mu\nu}$ are the first orders at which ${}^{(4)}\!g_{00}$ can appear. At the same time, $\varphi=O(\varepsilon^{2})$ and $\sqrt{-g}=1+O(\varepsilon^{2})$, so that the correction to $\sqrt{-g}$ would only generate $O(\varepsilon^{6})$ terms when multiplying the quadratic invariants.

Using the SPN expressions of Sec.~\ref{sec:PNformalism}, the linearized curvatures entering the 1PN action are
\begin{equation}\label{linearized-curvatures-for-Lextra}
{}^{(2)}\!R_{00}=\tfrac{1}{c^2}\nabla^2\Phi\,,\qquad
{}^{(2)}\!R_{ij}=\tfrac{1}{c^2}\nabla^2\Psi\,\delta_{ij} + \tfrac{1}{c^2}(\Psi-\Phi)_{,ij}\,,
\qquad
{}^{(2)}\!R=\frac{1}{c^2}\bigl(2\nabla^2\Phi-4\nabla^2\Psi\bigr)\,.
\end{equation}
Hence
\begin{equation}\label{quadratic-invariants-expansion}
R^2 = \bigl({}^{(2)}\!R\bigr)^2 + 2\,{}^{(2)}\!R\,{}^{(4)}\!R + O(\varepsilon^{8})\,,
\qquad
R_{\mu\nu}R^{\mu\nu}
=
{}^{(2)}\!R_{\mu\nu}{}^{(2)}\!R\,^{\mu\nu}
+
2\,{}^{(2)}\!R_{\mu\nu}{}^{(4)}\!R\,^{\mu\nu}
+ O(\varepsilon^{8})\,.
\end{equation}
Since ${}^{(2)}\!R=O(\varepsilon^{2})$ and ${}^{(4)}\!R=O(\varepsilon^{4})$, the mixed terms are $O(\varepsilon^{6})$ and therefore lie beyond 1PN accuracy. The same holds for the terms involving ${}^{(4)}\!R_{\mu\nu}$.

It follows that the extra-curvature part of the Lagrangian contributes at 1PN only through the linearized curvatures and scalar perturbation,
\begin{equation}\label{Lextra-1PN-structure}
L_{g}^{\,\textnormal{extra}}
=
\frac{c^4}{16\pi G} \int d^3\mathbf{x}\,
\biggl[
\frac{f_{RR}}{2}\,\bigl({}^{(2)}\!R\bigr)^2
+
f_Y\,{}^{(2)}\!R_{\mu\nu}{}^{(2)}\!R\,^{\mu\nu}
+
f_{R\phi}\,{}^{(2)}\!R\,\varphi
+
\frac{f_{\phi\phi}}{2}\,\varphi^2
-
\frac{1}{2}\,\partial_i\varphi\,\partial_i\varphi
\biggr]
+O(\varepsilon^{6})\,,
\end{equation}
where the time-derivative part of $\varphi_{;\alpha}\varphi^{;\alpha}$ is $O(\varepsilon^{6})$ and is therefore beyond the required accuracy. Equation~\eqref{Lextra-1PN-structure} shows directly that ${}^{(4)}\!g_{00}$ is absent from the genuine higher-curvature sector at 1PN order.

This conclusion is fully consistent with the modified Poisson equation for $\Phi$, namely
\begin{equation}\label{modified-Poisson}
(\nabla^2 - m_Y^2)\,\nabla^2\Phi + c^2 \,\biggl[\frac{m_Y^2}{2} - \frac{m_R^2 + 2m_Y^2}{6\,m_R^2}\,\nabla^2\biggr]{}^{(2)}\!R + c^2 \,m_Y^2\,f_{R\phi}\,\nabla^2\varphi = -m_Y^2\,\mathcal{X}\,c^4\sum_a m_a\,\delta(\mathbf{x} - \mathbf{x}_a)\,,
\end{equation}
where the right-hand side has been rewritten in terms of the Newtonian mass density associated with the point-particle source, after using $T_{00}=\rho\,c^2$ and the convention ${}^{(2)}\!g_{00}=2\Phi/c^2$. 

In other words, Eq.~\eqref{modified-Poisson} is the potential form of the first line of Eq.~\eqref{PMfieldequationFOG3}, expressed after factoring out the conventional powers of $c$ entering the definition of $\Phi$ and of the matter source in the weak-field expansion. This normalization makes the Poisson structure explicit and will be assumed throughout the remainder of the derivation, which can be rewritten as
\begin{equation}\label{Poisson-rearranged}
\nabla^2\Phi = \mathcal{X}\,c^4 \sum_a m_a\,\delta(\mathbf{x} - \mathbf{x}_a) + \mathcal{F}(\Phi,{}^{(2)}\!R,\varphi)\,,
\end{equation}
with
\begin{equation}\label{F-correction}
\mathcal{F} \equiv \frac{\nabla^4\Phi}{m_Y^2} + \frac{c^2}{2}\,{}^{(2)}\!R - \frac{c^2 (m_R^2 + 2m_Y^2)}{6\,m_R^2\,m_Y^2}\,\nabla^2{}^{(2)}\!R + c^2 f_{R\phi}\,\nabla^2\varphi\,.
\end{equation}
In the Einstein--Hilbert sector \eqref{Sg-def}, this converts the field contribution with ${}^{(4)}\!g_{00}$ into
\begin{equation}\label{field-sector-h4}
\frac{c^2}{16\pi G} \int\! {}^{(4)}\!g_{00}\,\nabla^2\Phi \,d^3\mathbf{x}
=
\int\!\biggl[\frac{1}{2}\,{}^{(4)}\!g_{00} \,c^2 \sum_a m_a \delta(\mathbf{x}-\mathbf{x}_a)\biggr]\,d^3\mathbf{x}
+
\frac{c^2}{16\pi G} \int\!{}^{(4)}\!g_{00}\,\mathcal{F}\,d^3\mathbf{x}\,.
\end{equation}
The first term yields the same ${}^{(4)}\!g_{00}$ cancellation as in GR. Any residual contribution involving the second term would scale as
\begin{equation}\label{order-estimate}
{}^{(4)}\!g_{00} \times \mathcal{F}
\sim
O(\varepsilon^{4}) \times O(\varepsilon^{2})
=
O(\varepsilon^{6})\,,
\end{equation}
and is therefore beyond the 1PN accuracy of the Lagrangian.

The total 1PN Lagrangian of STFOG is therefore independent of ${}^{(4)}\!g_{00}$: in the Einstein--Hilbert plus matter sectors through an exact cancellation, and in the genuinely extra-curvature sector through post-Newtonian order counting. Hence the linearized metric \eqref{ds2-PN} is both necessary and sufficient for the 1PN dynamics of the relativistic $N$-body system in STFOG.

\subsection{STFOG Post-Newtonian \texorpdfstring{$N$}{N}-body Lagrangian}\label{sec:N-body-Lagrangian}

Having established that only the linearized potentials $\Phi$, $A_i$, $X$ enter the 1PN Lagrangian, we insert the STFOG solutions from Sec.~\ref{sec:linearized_field_equations} into the variational Lagrangian \eqref{variational-Lagrangian}. Under the hypothesis $\Phi \sim \Psi$ (i.e.\ $\gamma \sim 1$, cf.\ Sec.~\ref{sec:brumberg}), with $\mathbf{v}_a = d\mathbf{x}_a/dt$, $\mathbf{x}_{ab} = \mathbf{x}_a - \mathbf{x}_b$, $r_{ab} = |\mathbf{x}_{ab}|$, and $\mathbf{n}_{ab} = \mathbf{x}_{ab}/r_{ab}$\footnote{All quantities are evaluated at the same coordinate time $t$.}, the total Lagrangian for the orbital motion of the $a$-th body in the field of the $N-1$ others is given by Eq.~\eqref{variational-Lagrangian}, 
where $\Phi^{(a)} = \sum_{b\neq a}\Phi_b(\mathbf{x}_a)$, $A_i^{(a)} = \sum_{b\neq a}(A_i)_b(\mathbf{x}_a)$, $X^{(a)} = \sum_{b\neq a}X_b(\mathbf{x}_a)$, with the individual potentials given by Eqs.~\eqref{Phi-point}, \eqref{Ai-point}, and \eqref{Superpotential-result}. 

By inserting the post-Newtonian potentials $\Phi, \Psi$ and $A_i$ and the superpotential $X$ that provide the correct forces on each body, and arranging the body labels, we find the STFOG Lagrangian for the relativistic $N$-body system
\begin{eqnarray}\label{Final-relativistic-lagrangian}
L &=& \sum_{a=1}^{N}\frac{m_{a}\,v_{a}^2}{2}\biggl(1 + 3\sum_{b\ne a}\frac{G\,m_{b}}{c^2\,r_{ab}}\bigl[1 + \zeta(r_{ab})\bigr]\biggr) + \sum_{a=1}^{N}\frac{m_{a}\,v_{a}^4}{8\,c^2} + \sum_{a=1}^{N}\sum_{b\ne a}\frac{G\,m_{a}\,m_{b}}{2\,r_{ab}}\bigl[1 + \zeta(r_{ab})\bigr]
\nonumber\\[6pt]
&&-\;\sum_{a=1}^{N}\sum_{b\ne a}\frac{G\,m_{a}\,m_{b}}{4\,c^2\,r_{ab}}\bigl[7\,\mathbf{v}_{a}\!\cdot\!\mathbf{v}_{b} + (\mathbf{v}_{a}\!\cdot\!\mathbf{n}_{ab})(\mathbf{v}_{b}\!\cdot\!\mathbf{n}_{ab})\bigr] - \sum_{a=1}^{N}\sum_{b\ne a}\frac{G\,m_{a}\,m_{b}}{2\,c^2}\;\mathcal{W}(r_{ab})
\nonumber\\[6pt]
&&-\;\sum_{a=1}^{N}\sum_{b\ne a}\sum_{c\ne a}\frac{G^2\,m_{a}\,m_{b}\,m_{c}}{2\,c^2\,r_{ab}\,r_{ac}}\;\Xi(r_{ab},r_{ac})\,,
\end{eqnarray}
where $\mathbf{n}_{ab}$ is the unit separation vector.
The first line of \eqref{Final-relativistic-lagrangian} contains the kinetic energy, the 1PN kinetic potential coupling (proportional to $v_a^2\,\Phi$), the $v^4/c^2$ correction, and the Newtonian potential energy modified by $\zeta(r_{ab})$ (Eq.~\eqref{zeta-explicit}). The second line includes the GR gravitomagnetic coupling and the STFOG gravitomagnetic extra-function $\mathcal{W}$, which encodes the Yukawa corrections from $A_i$ and $X_{,0i}$. The third line is the three-body potential $\Xi$, arising from $(\Phi^{(a)})^2$.

\subsubsection{Yukawa gravitomagnetic function \texorpdfstring{$\mathcal{W}(r_{ab})$}{W(rab)}}

The function $\mathcal{W}(r_{ab})$ collects the contributions from the vector potential $A_i$ and the superpotential derivatives $X_{,0i}$ in $Z_i = A_i - X_{,0i}$. To write it compactly, we introduce for each Yukawa mode $i \in \{\pm,Y\}$ with mass $m_i$ the auxiliary functions

\begin{equation}\label{auxiliary-functions}
\mathcal{P}_i(r) \equiv \frac{1 - (1 + m_i\,r)\,e^{-m_i\,r}}{m_i^2\,r^3}\,,\qquad
\mathcal{Q}_i(r) \equiv \biggl[\frac{1}{r} + \frac{2\,(1 + m_i\,r)}{m_i^2\,r^3}\biggr]e^{-m_i\,r} - \frac{2}{m_i^2\,r^3}\,.
\end{equation}

In terms of these, and denoting $F_+ \equiv g(\xi,\eta)$, $F_- \equiv \tfrac{1}{3} - g(\xi,\eta)$, $F_Y \equiv -\tfrac{4}{3}$, the gravitomagnetic extra-function is
\begin{equation}\label{W-function}
\mathcal{W}(r_{ab}) = -\frac{4\,e^{-m_Y r_{ab}}}{r_{ab}}\;(\mathbf{v}_a\!\cdot\!\mathbf{v}_b) + \sum_{i=\pm,Y}(-F_i)\,\biggl\{\bigl[(\mathbf{v}_a\!\cdot\!\mathbf{v}_b) - (\mathbf{v}_a\!\cdot\!\mathbf{n}_{ab})(\mathbf{v}_b\!\cdot\!\mathbf{n}_{ab})\bigr]\,\mathcal{P}_i(r_{ab}) + (\mathbf{v}_a\!\cdot\!\mathbf{n}_{ab})(\mathbf{v}_b\!\cdot\!\mathbf{n}_{ab})\;\mathcal{Q}_i(r_{ab})\biggr\}\,.
\end{equation}
The first term is the Yukawa correction to the gravitomagnetic vector potential $A_i$ (induced by the $R_{\alpha\beta}R^{\alpha\beta}$ invariant), while the remaining terms arise from the superpotential derivatives $X_{,0i}$, with each mode $i$ contributing through the coupling strengths $F_i$. In the GR limit ($m_i \to \infty$ for all $i$), one has $\mathcal{P}_i \to 0$ and $\mathcal{Q}_i \to 0$, so that $\mathcal{W} \to 0$ and the standard EIH Lagrangian is recovered.

\subsubsection{Three-body Yukawa function \texorpdfstring{$\Xi(r_{ab},r_{ac})$}{Xi(rab,rac)}}

The three-body potential function arises from $(\Phi^{(a)})^2$ and reads
\begin{equation}\label{Xi-function}
\Xi(r_{ab},r_{ac}) = \bigl[1 + \zeta(r_{ab})\bigr]\,\bigl[1 + \zeta(r_{ac})\bigr]\,,
\end{equation}
which, upon expanding with $\zeta(r) = \sum_i F_i\,e^{-m_i r}$ ($i = \pm,Y$), generates nine cross-terms of the form $F_i\,F_j\,e^{-m_i r_{ab} - m_j r_{ac}}$ in addition to the linear Yukawa terms. In GR ($\zeta \to 0$), one recovers $\Xi \to 1$.

\subsection{Euler--Lagrange orbital equations}\label{sec:EOM}

The post-Newtonian equations of motion for the $a$-th body follow from the Euler--Lagrange equations
\begin{equation}\label{Euler-Lagrange-equations}
\frac{d}{dt}\,\frac{\partial L}{\partial\mathbf{v}_a} - \frac{\partial L}{\partial\mathbf{x}_a} = 0\,,\qquad a = 1,2,\ldots,N\,,
\end{equation}
together with the Lagrangian \eqref{Final-relativistic-lagrangian}. These provide the most general system of relativistic differential equations governing the orbital motion of an isolated $N$-body system in the STFOG class, i.e. the geodesics in the effective 1PN metric generated by the other $N-1$ bodies. 
By construction, they reduce to the Einstein--Infeld--Hoffmann equations \cite{EIH,Brumberg2,Soffel} in the GR limit. The regular STFOG subclasses are immediately recovered by specializing the parameters: $g(\xi,\eta)=1/3$ for $f(R,R_{\alpha\beta}R^{\alpha\beta})$ and $m_Y\to\infty$ (absence of the $R_{\alpha\beta}R^{\alpha\beta}$ sector) for $f(R,\phi)$. The NCSG case, instead, is obtained as the regular $m_R\to\infty$ (scalaron-decoupling) limit discussed in Secs.~\ref{sec:NCSG-solutions} and \ref{sec:NCSG-EOM}, where the scalar modes decouple and the single surviving beyond-GR scale is $\beta=m_Y$.

\subsubsection{STFOG Post-Newtonian N-body equations of motion}
We now cast the Euler--Lagrange system \eqref{Euler-Lagrange-equations} in explicit acceleration form. Splitting Eq.~\eqref{Final-relativistic-lagrangian} as
\(L=L_{\rm N}+c^{-2}L_{\rm PN}\), with
\[
L_{\rm N}
=
\sum_{a=1}^{N}\frac{m_a v_a^2}{2}
+
\frac{1}{2}\sum_{a=1}^{N}\sum_{b\neq a}
\frac{Gm_a m_b}{r_{ab}}\bigl[1+\zeta(r_{ab})\bigr]\,,
\]
the Newtonian-order STFOG acceleration is
\begin{equation}\label{STFOG-Newtonian-acceleration}
\frac{d^{2}\mathbf{x}^{\rm N}_{a}}{dt^{2}}
=
-\sum_{b\neq a}
\frac{Gm_b}{r_{ab}^{2}}\,\mathbf{n}_{ab}
\left[
1+\zeta(r_{ab})-r_{ab}\zeta'(r_{ab})
\right]\,.
\end{equation}
The complete first post-Newtonian equations of motion are then obtained as
\begin{equation}\label{STFOG-compact-EOM}
\frac{d^{2}\mathbf{x}_{a}}{dt^{2}}
=
\frac{d^{2}\mathbf{x}^{\rm N}_{a}}{dt^{2}}
+
\frac{1}{m_a c^{2}}
\left[
\frac{\partial L_{\rm PN}}{\partial \mathbf{x}_{a}}
-
\frac{d}{dt}
\left(
\frac{\partial L_{\rm PN}}{\partial \mathbf{v}_{a}}
\right)
\right],
\qquad
a=1,\ldots,N\,.
\end{equation}
where \(L_{\rm PN}\) is read off from Eq.~\eqref{Final-relativistic-lagrangian}.
In the last term of this equation, all second-time-derivative terms generated at intermediate steps are consistently reduced by means of the Newtonian-order STFOG equation of motion. Here the Newtonian-order acceleration entering the order-reduced 1PN terms is
\begin{equation}\label{STFOG-Newtonian-complete}
\frac{d^2\mathbf x_a^{\rm N}}{dt^2}
=
-\sum_{b\neq a}
\frac{Gm_b}{r_{ab}^{2}}\,\mathbf n_{ab}
\left[
1+\zeta(r_{ab})-r_{ab}\zeta'(r_{ab})
\right],
\qquad a=1,\ldots,N .
\end{equation}
The scalar Yukawa function and its radial derivative are
\begin{equation}\label{zeta-EOM}
\zeta(r)=\sum_{i=\pm,Y}F_i e^{-m_i r},
\qquad
\zeta'(r)=-\sum_{i=\pm,Y}F_i m_i e^{-m_i r},
\qquad
F_+=g(\xi,\eta),
\qquad
F_-=\frac{1}{3}-g(\xi,\eta),
\qquad
F_Y=-\frac{4}{3}.
\end{equation}
We shall only use the relative velocity $\mathbf v_{ab}\equiv\mathbf v_a-\mathbf v_b$. For each Yukawa mode, the auxiliary functions generated by the SPN superpotential contribution are those defined in Eq. \eqref{auxiliary-functions}; for the explicit acceleration formula below, one also needs their radial derivatives,
\begin{equation}\label{P-Q-prime-functions-EOM}
\mathcal P_i'(r)
=
\frac{
\left(m_i^2r^2+3m_i r+3\right)e^{-m_i r}-3
}{m_i^2r^4},
\qquad
\mathcal Q_i'(r)
=
\frac{
6-\left(m_i^3r^3+3m_i^2r^2+6m_i r+6\right)e^{-m_i r}
}{m_i^2r^4}.
\end{equation}
All 1PN contributions obtained from Eqs.~\eqref{Final-relativistic-lagrangian} and \eqref{STFOG-compact-EOM} can then be collected into the final equation
\begin{equation}\label{STFOG-single-EOM}
\begin{aligned}
\frac{d^2\mathbf x_a}{dt^2}
={}&
\frac{d^2\mathbf x_a^{\rm N}}{dt^2}
+
\frac{1}{c^2}
\biggl\{
-
\left(
\mathbf v_a\cdot\frac{d^2\mathbf x_a^{\rm N}}{dt^2}
\right)\mathbf v_a
-
\frac{1}{2}v_a^2\frac{d^2\mathbf x_a^{\rm N}}{dt^2}
\\[2mm]
&+
\frac{3}{2}
\sum_{b\neq a}Gm_b
\left(v_a^2+v_b^2\right)
\frac{r_{ab}\zeta'(r_{ab})-1-\zeta(r_{ab})}{r_{ab}^{2}}\,
\mathbf n_{ab}
-
3
\left[
\sum_{c\neq a}
\frac{Gm_c}{r_{ac}}\bigl(1+\zeta(r_{ac})\bigr)
\right]
\frac{d^2\mathbf x_a^{\rm N}}{dt^2}
\\[2mm]
&-
3\sum_{b\neq a}Gm_b
\frac{r_{ab}\zeta'(r_{ab})-1-\zeta(r_{ab})}{r_{ab}^{2}}
\left(\mathbf n_{ab}\cdot\mathbf v_{ab}\right)\mathbf v_a
\\[2mm]
&+
\frac{1}{2}\sum_{b\neq a}
\frac{Gm_b}{r_{ab}^{2}}
\biggl[
\left(
7\mathbf v_a\cdot\mathbf v_b
+
3(\mathbf v_a\cdot\mathbf n_{ab})(\mathbf v_b\cdot\mathbf n_{ab})
\right)\mathbf n_{ab}
-
(\mathbf v_b\cdot\mathbf n_{ab})\mathbf v_a
-
(\mathbf v_a\cdot\mathbf n_{ab})\mathbf v_b
\biggr]
\\[2mm]
&+
\frac{1}{2}\sum_{b\neq a}Gm_b
\biggl[
-
\frac{\mathbf n_{ab}\cdot\mathbf v_{ab}}{r_{ab}^{2}}
\left(
7\mathbf v_b+(\mathbf v_b\cdot\mathbf n_{ab})\mathbf n_{ab}
\right)
\\[1mm]
&\hspace{2.3cm}
+
\frac{1}{r_{ab}}
\biggl(
7\frac{d^2\mathbf x_b^{\rm N}}{dt^2}
+
\left[
\frac{d^2\mathbf x_b^{\rm N}}{dt^2}\cdot\mathbf n_{ab}
+
\frac{
\mathbf v_a\cdot\mathbf v_b-v_b^2-(\mathbf n_{ab}\cdot\mathbf v_{ab})(\mathbf v_b\cdot\mathbf n_{ab})
}{r_{ab}}
\right]
\mathbf n_{ab}
\\[1mm]
&\hspace{3.4cm}
+
\frac{\mathbf v_b\cdot\mathbf n_{ab}}{r_{ab}}
\left[
\mathbf v_{ab}
-
(\mathbf n_{ab}\cdot\mathbf v_{ab})\mathbf n_{ab}
\right]
\biggr)
\biggr]
\\[2mm]
&+
\sum_{b\neq a}Gm_b
\biggl\{
4\frac{(1+m_Yr_{ab})e^{-m_Yr_{ab}}}{r_{ab}^{2}}
(\mathbf n_{ab}\cdot\mathbf v_{ab})\mathbf v_b
-
4\frac{e^{-m_Yr_{ab}}}{r_{ab}}
\frac{d^2\mathbf x_b^{\rm N}}{dt^2}
\\[1mm]
&\hspace{2.4cm}
-
4(\mathbf v_a\cdot\mathbf v_b)
\frac{(1+m_Yr_{ab})e^{-m_Yr_{ab}}}{r_{ab}^{2}}
\mathbf n_{ab}
\\[1mm]
&\hspace{2.4cm}
+
\sum_{i=\pm,Y}(-F_i)
\biggl[
\mathcal P_i(r_{ab})
\frac{d^2\mathbf x_b^{\rm N}}{dt^2}
+
\mathcal P_i'(r_{ab})
(\mathbf n_{ab}\cdot\mathbf v_{ab})\mathbf v_b
\\[1mm]
&\hspace{3.1cm}
+
\left(
\frac{d^2\mathbf x_b^{\rm N}}{dt^2}\cdot\mathbf n_{ab}
+
\frac{
\mathbf v_a\cdot\mathbf v_b-v_b^2-(\mathbf n_{ab}\cdot\mathbf v_{ab})(\mathbf v_b\cdot\mathbf n_{ab})
}{r_{ab}}
\right)
\left[
\mathcal Q_i(r_{ab})-
\mathcal P_i(r_{ab})
\right]
\mathbf n_{ab}
\\[1mm]
&\hspace{3.1cm}
+
(\mathbf v_b\cdot\mathbf n_{ab})
\left[
\mathcal Q_i'(r_{ab})-\mathcal P_i'(r_{ab})
\right]
(\mathbf n_{ab}\cdot\mathbf v_{ab})\mathbf n_{ab}
\\[1mm]
&\hspace{3.1cm}
+
(\mathbf v_b\cdot\mathbf n_{ab})
\left[
\mathcal Q_i(r_{ab})-\mathcal P_i(r_{ab})
\right]
\frac{\mathbf v_{ab}-(\mathbf n_{ab}\cdot\mathbf v_{ab})\mathbf n_{ab}}{r_{ab}}
\\[1mm]
&\hspace{3.1cm}
-
\left[
(\mathbf v_a\cdot\mathbf v_b)\mathcal P_i'(r_{ab})
+
(\mathbf v_a\cdot\mathbf n_{ab})(\mathbf v_b\cdot\mathbf n_{ab})
\left(
\mathcal Q_i'(r_{ab})-
\mathcal P_i'(r_{ab})
\right)
\right]
\mathbf n_{ab}
\\[1mm]
&\hspace{3.1cm}
-
\frac{\mathcal Q_i(r_{ab})-\mathcal P_i(r_{ab})}{r_{ab}}
\left[
(\mathbf v_b\cdot\mathbf n_{ab})\mathbf v_a
+
(\mathbf v_a\cdot\mathbf n_{ab})\mathbf v_b
-
2(\mathbf v_a\cdot\mathbf n_{ab})(\mathbf v_b\cdot\mathbf n_{ab})\mathbf n_{ab}
\right]
\biggr]
\biggr\}
\\[2mm]
&-
\sum_{b\neq a}Gm_b
\frac{r_{ab}\zeta'(r_{ab})-1-\zeta(r_{ab})}{r_{ab}^{2}}
\left[
\sum_{c\neq a}
\frac{Gm_c}{r_{ac}}\bigl(1+\zeta(r_{ac})\bigr)
+
\sum_{c\neq b}
\frac{Gm_c}{r_{bc}}\bigl(1+\zeta(r_{bc})\bigr)
\right]
\mathbf n_{ab}
\biggr\} .
\end{aligned}
\end{equation}
Equation~\eqref{STFOG-single-EOM} is order-reduced: every acceleration appearing on its right-hand side is the Newtonian STFOG acceleration \eqref{STFOG-Newtonian-complete}. The first line inside braces comes from the special-relativistic kinetic correction, the next two lines from the kinetic-potential coupling, the following two blocks from the ordinary EIH velocity-dependent term, the long Yukawa block from the gravitomagnetic function $\mathcal W$, and the last line from the three-body potential generated by $(\Phi^{(a)})^2$. Therefore no derivative of the three-body function $\Xi$ is omitted; its contribution is already contained in the last line through the two Newtonian potentials centered on bodies $a$ and $b$.

\subsubsection{NCSG specialization}\label{sec:NCSG-EOM}

For the NCSG sector one has
\begin{equation}\label{NCSG-parameters-EOM}
\zeta(r)
=
-\frac{4}{3}e^{-\beta r},
\qquad
\zeta'(r)
=
\frac{4\beta}{3}e^{-\beta r},
\qquad
m_R\to\infty,
\qquad
m_Y=\beta,
\qquad
F_Y=-\frac{4}{3},
\qquad
F_+=F_-=0 .
\end{equation}
and the Newtonian-order NCSG acceleration becomes
\begin{equation}\label{NCSG-Newtonian-acceleration}
\frac{d^2\mathbf x_a^{\rm N}}{dt^2}
=
-
\sum_{b\neq a}
\frac{Gm_b}{r_{ab}^{2}}\mathbf n_{ab}
\left[
1-\frac{4}{3}(1+\beta r_{ab})e^{-\beta r_{ab}}
\right].
\end{equation}
The complete NCSG 1PN $N$-body equations of motion are obtained directly from Eq.~\eqref{STFOG-single-EOM} by the substitutions \eqref{NCSG-parameters-EOM}, with $\mathcal P_i,\mathcal Q_i$ replaced by $\mathcal P_\beta,\mathcal Q_\beta$. Equivalently, the surviving gravitomagnetic Yukawa function is
\begin{equation}\label{NCSG-W-final}
\mathcal W_{ab}
=
-
\frac{4e^{-\beta r_{ab}}}{r_{ab}}
(\mathbf v_a\cdot\mathbf v_b)
+
\frac{4}{3}
\left\{
\left[
\mathbf v_a\cdot\mathbf v_b
-
(\mathbf v_a\cdot\mathbf n_{ab})(\mathbf v_b\cdot\mathbf n_{ab})
\right]
\mathcal P_\beta(r_{ab})
+
(\mathbf v_a\cdot\mathbf n_{ab})(\mathbf v_b\cdot\mathbf n_{ab})
\mathcal Q_\beta(r_{ab})
\right\} .
\end{equation}
The first term in $\mathcal W_{ab}$ is the genuine Ricci-tensor Yukawa correction to the vector potential, while the $\mathcal P_\beta$ and $\mathcal Q_\beta$ terms come from the SPN superpotential contribution $X_{,0i}$. In the limit $\beta\to\infty$, one has $\zeta_\beta\to0$, $\mathcal W_{ab}\to0$, and the standard Einstein--Infeld--Hoffmann equations are recovered.

%------------------------------------------------------
\subsection{Astrophysical regimes and observational tests}\label{sec:dynamics-discussion}

The Lagrangian \eqref{Final-relativistic-lagrangian}, the compact Euler--Lagrange relation \eqref{Euler-Lagrange-equations}--\eqref{STFOG-compact-EOM}, and the final equation \eqref{STFOG-single-EOM} constitute the main result of the present work: they provide the complete 1PN equations of orbital motion for the relativistic $N$-body system within the regular STFOG branch. The ETG corrections are encoded in the Yukawa functions $\zeta$, $\mathcal{W}$ and $\Xi$, parameterized by the effective masses $(m_R,\,m_Y,\,m_\phi)$ and the coupling parameters $(\xi,\,\eta,\,g(\xi,\eta))$. Specializing these parameters one immediately recovers the subclasses $f(R)$, $f(R,\phi)$ and $f(R,R_{\alpha\beta}R^{\alpha\beta})$. The NCSG sector, by contrast, is obtained as the $m_R\to\infty$ (scalaron-decoupling) limit of Secs.~\ref{sec:NCSG-solutions} and \ref{sec:NCSG-EOM}, where the scalar modes decouple and $\beta$ is the only surviving beyond-GR mass scale.

The formalism covers any astrophysical system in the 1PN regime. Binary pulsars such as PSR~J0737--3039 \cite{Brumberg3,Kramer2021} are a natural arena: the two neutron-star masses are comparable ($m_1 \sim 1.34\,M_\odot$, $m_2 \sim 1.25\,M_\odot$), so that the $N = 2$ specialization of \eqref{Final-relativistic-lagrangian} yields the relativistic periastron advance \cite{DamourDeruelle} and a setup for high-precision gravitational tests in which the explicit Yukawa corrections can be confronted with pulsar-timing data. For stellar orbits around Sgr\,A$^*$, S2 \cite{GRAVITY2020} being the paradigmatic example, the equations supply the orbital precession as modified by the ETG effective masses, thereby extending the constraints of Table~\ref{tab:bounds-STFOG-NCSG}. Hierarchical and resonant triple stellar systems can be analyzed through the three-body terms $\Xi(r_{ab},r_{ac})$, which capture the nonlinear gravitational coupling enhanced (or suppressed) by the Yukawa interactions.

Galactic-center astrometry deserves a dedicated discussion, as it shows concretely how the closed-form results obtained here --- the $N$-body Lagrangian \eqref{Final-relativistic-lagrangian} and the equations of motion \eqref{STFOG-single-EOM} --- can sharpen the estimates of extended-gravity parameters. The GRAVITY detection of the Schwarzschild precession of S2 ($f_{\rm SP} = 1.10 \pm 0.19$ \cite{GRAVITY2020}), refined by later multi-star determinations \cite{GRAVITY2022}, has already constrained several extended models --- power-law $f(R)$, generic Yukawa corrections, scalar-tensor and non-local gravity \cite{Borka2021} --- and lowered the graviton-mass bound to $m_g \lesssim 1.8\times10^{-22}\,\mathrm{eV}$ from S2 alone, with forecasted gains up to $\sim 15$ as the precession approaches its GR value \cite{Jovanovic2024}. All these analyses rest on Yukawa potentials of the form \eqref{Phi-Yukawa}: their coupling--range pair $(\delta,\Lambda)$ maps, up to a normalization of the effective coupling, onto our $(F_i,\,m_i^{-1})$, so that the short-range branch ($m_i\,r \gg 1$) of Table~\ref{tab:bounds-STFOG-NCSG} and the long-range branch ($m_i\,r \ll 1$) relevant for graviton-mass bounds are complementary sides of a single exclusion region, both in the small-deviation regime $\Phi\sim\Psi$ where our construction holds.

The relativistic 1PN $N$-body dynamics derived here sharpens these estimates on three fronts. The first is self-consistency: whereas Refs.~\cite{Borka2021,Jovanovic2024} superpose a Yukawa-modified Newtonian force on the general-relativistic 1PN correction, in Eq.~\eqref{STFOG-single-EOM} the same parameters $(F_i,m_i)$ drive the \emph{full} 1PN dynamics --- through $\zeta$ and $\zeta'$, the gravitomagnetic functions $\mathcal{P}_i,\mathcal{Q}_i$ in $\mathcal{W}$, and the three-body term $\Xi$ --- so that the modelled precession is complete at the retained order and the posteriors on $(F_i,m_i)$ are shielded from the offsets of a Newtonian-only modification. The second is the genuinely $N$-body character of the description: perturbers such as the other S-stars, an extended mass enclosed within the S2 orbit, or a companion of Sgr\,A$^*$ --- which mimic part of the beyond-GR signal through a retrograde precession --- are evolved in the same dynamics as the theory parameters rather than added by hand, a degeneracy control that becomes decisive at the current astrometric accuracy of the GRAVITY's interferometer \cite{GRAVITY2022}. The third is the possibility of joint, mode-resolved fits: since Eq.~\eqref{STFOG-single-EOM} holds for arbitrary $N$ with the three mass scales $(m_R,m_Y,m_\phi)$ kept distinct, a single parameter set can be confronted with several orbits at once (S2, S29, S38, S55, \dots), the very strategy expected to push graviton-mass bounds towards $10^{-24}\,\mathrm{eV}$ as the precession approaches its GR value \cite{Jovanovic2024}. The results \eqref{Final-relativistic-lagrangian} and \eqref{STFOG-single-EOM} thus supply the dynamical input needed to turn forthcoming Galactic-center astrometry into sharper, bias-controlled constraints on the STFOG parameter space $(F_\pm,F_Y;\,m_\pm,m_Y)$ and on the NCSG scale $\beta$. 

We leave these applications to future work: the natural next step is to extract observable timing and precession formulas from the general Lagrangian, so as to confront them with high-precision pulsar-timing data, long-term monitoring of Galactic-center stellar orbits such as S2 around Sgr\,A$^*$, and future observations of multi-body relativistic systems.

\section{Conclusions}\label{sec:conclusions}

We have derived the complete first post-Newtonian (1PN) $N$-body dynamics of the regular Scalar-Tensor-Fourth-Order Gravity class in the screened regime $(\Phi\sim\Psi)$. Starting from the STFOG action, in the fixed Standard Post-Newtonian (SPN) gauge, we solved the linearized fourth-order field equations by the Green-function method for a source given by a system of $N$ bodies, and inserted the resulting potentials into the point-particle variational principle. In the adopted SPN gauge, we determined the set of post-Newtonian gravitational potentials \eqref{Phi-point}--\eqref{Psi-point}--\eqref{Zi-equation}--\eqref{Zi-Nbody} including the full gravitomagnetic sector and the NCSG limit \eqref{NCSG-potentials}--\eqref{NCSG-Zi}, in agreement with the corresponding harmonic-gauge results of Refs.~\cite{FOG_CGL,PRD1,Lambiase:2013dai,NaefJetzer2010,FOG1,FOG2,FOGST}, once the standard SPN--harmonic transformation is taken into account. In particular, in the NCSG sector we obtained the linearized metric generated by a source consisting of $N$ material point masses. Through the variational principle, we then find the closed-form 1PN $N$-body Lagrangian \eqref{Final-relativistic-lagrangian} for STFOG and the corresponding orbital equations.

The key analytic results are the Lagrangian \eqref{Final-relativistic-lagrangian}, the Yukawa functions $\zeta$, $\mathcal W$, and $\Xi$ defined through \eqref{zeta-explicit}, \eqref{W-function}, and \eqref{Xi-function}, the compact Euler--Lagrange system \eqref{Euler-Lagrange-equations}--\eqref{STFOG-compact-EOM}, and the complete explicit equations of motion collected in \eqref{STFOG-single-EOM}. This closed form includes the full variations of the Yukawa structures, namely the derivatives of $\zeta$, $\mathcal P_i$, $\mathcal Q_i$, and the complete variation of the three-body function $\Xi$. Therefore no Yukawa contribution is left implicit in the final orbital dynamics.

A result of conceptual relevance is the extension of the Brumberg argument to the STFOG class in the screened regime $\Phi\sim\Psi$ $(\gamma\sim1)$, enforced by the chameleon mechanism in the scalar sector. The nonlinear component ${}^{(4)}\!g_{00}$ is irrelevant for the 1PN orbital dynamics for two separate reasons: in the Einstein--Hilbert plus matter sectors it disappears through the exact Brumberg cancellation, while in the genuinely higher-curvature sector it is excluded by post-Newtonian order counting. The linearized metric is therefore sufficient to determine the complete 1PN equations of motion in the full regular STFOG branch. 

The principal subclasses follow directly from the general result. The $f(R)$, $f(R,\phi)$, and $f(R,R_{\alpha\beta}R^{\alpha\beta})$ sectors are obtained by the appropriate limits of the effective masses and coupling coefficients $(m_R,m_Y,m_\phi;\xi,\eta,g)$. The Non-Commutative Spectral Geometry sector corresponds to the scalaron-decoupling limit $m_R\to\infty$, with $m_Y=\beta$ as the only surviving beyond-GR mass scale and with the SPN gravitomagnetic completion retained explicitly. In the simultaneous limit in which all Yukawa corrections vanish, the Lagrangian and the post-Newtonian equations of motion reduce to the standard Einstein--Infeld--Hoffmann dynamics.

The formalism provides a common framework for the relativistic celestial mechanics beyond General Relativity in the context of Scalar-Tensor-Fourth-Order Gravity (STFOG), and applies to relativistic systems in which 1PN effects are measurable, while radiation reaction, explicit 2PN corrections and strong-field matching remain subdominant. Solar-System dynamics, binary pulsars such as PSR~J0737--3039, stellar orbits around Sgr\,A$^{*}$, and (hierarchical) triple systems are natural testing grounds for applications. A natural continuation of this work is the extraction of timing, precession and secular-evolution observables from the general STFOG Lagrangian, and their comparison with high-precision data from pulsar timing, Galactic-center astrometry and future multi-body relativistic systems. In particular, as discussed in Sec.~\ref{sec:dynamics-discussion}, the complete equations of motion \eqref{STFOG-single-EOM} supply the self-consistent 1PN dynamical input required to refine the estimates of extended-gravity parameters and the graviton-mass bounds currently inferred from the stellar orbits around Sgr\,A$^*$ \cite{Borka2021,GRAVITY2022,Jovanovic2024}.

\begin{acknowledgments}
The author thanks Antonio Capolupo, Gaetano Lambiase, Salvatore Capozziello and Sante Carloni for discussions and comments on related aspects of extended theories of gravity. The author also acknowledges the INFN QGSKY initiative and the Enrico Fermi Research Centre (CREF) for support.
\end{acknowledgments}

\end{document}